\documentclass[fleqn,usenatbib]{mnras}

\usepackage{newtxtext,newtxmath}

\usepackage[T1]{fontenc}
\usepackage[normalem]{ulem}

\DeclareRobustCommand{\VAN}[3]{#2}
\let\VANthebibliography\thebibliography
\def\thebibliography{\DeclareRobustCommand{\VAN}[3]{##3}\VANthebibliography}

\usepackage{graphicx}
\usepackage{amsmath}
\usepackage{example}
\usepackage{amsfonts}
\usepackage[section]{placeins}
\usepackage{caption}
\usepackage{subcaption}
\usepackage{hyperref}
\usepackage{import}
\usepackage{float}
\usepackage{siunitx}

\DeclareSIUnit\angstrom{\text{\AA}}

\newcommand{\md}          {{\rm d}}

\newcommand{\percc}       {{\rm cm}^{-3}}

\newcommand{\HII}          {H{\,\textsc{ii}}}

\newcommand{\K}           {\rm{K}}

\newcommand{\Myr}         {\rm{Myr}}
\newcommand{\Gyr}         {\rm{Gyr}}
\newcommand{\Msun}        {\rm{M}_\odot}

\newcommand{\Mpc}         {\rm{Mpc}}
\newcommand{\cMpc}        {\rm{cMpc}}

\newcommand{\kpc}         {\rm{kpc}}
\newcommand{\ckpc}        {\rm{ckpc}}
\newcommand{\pkpc}        {\rm{pkpc}}

\newcommand{\eV}          {\rm{eV}}
\newcommand{\erg}          {\rm{erg}}

\title[Simulating high-$z$ galaxies with a variable IMF]{Cosmological simulations of the high-redshift galaxy population adopting a variable stellar initial mass function}

\author[A. Durrant et al.]{Anna Durrant, $^{1}$\thanks{E-mail: A.R.Durrant@2023.LJMU.ac.uk }
Robert A. Crain,$^{1}$
Cedric G. Lacey,$^{2}$
Joop Schaye,$^{3}$
Renske Smit,$^{1}$
Andrea Gebek,$^{4}$
\newauthor
Matthieu Schaller,$^{3,5}$
Shengdong Lu,$^{2}$
Evgenii Chaikin,$^{2,3}$
Nick Andreadis,$^{4}$
Maarten Baes,$^{4}$
\newauthor
Matthew R. Bate,$^{6}$
Alejandro Ben\'itez-Llambay,$^{7}$
Carlos S. Frenk,$^{2}$
Filip Hu{\v{s}}ko,$^{3}$
Robert J. McGibbon,$^{3}$
\newauthor
Sylvia Ploeckinger,$^{8}$
and Alexander J. Richings$^{9,10}$
\\
\\
$^{1}$Astrophysics Research Institute, Liverpool John Moores University, 146 Brownlow Hill, Liverpool L3 5RF, UK\\
$^{2}$Institute for Computational Cosmology, Department of Physics, University of Durham, South Road, Durham, DH1 3LE, UK\\
$^{3}$Leiden Observatory, Leiden University, PO Box 9513, 2300 RA Leiden, the Netherlands\\
$^{4}$Department of Physics and Astronomy, Universiteit Gent, Proeftuinstraat 86 N3, B-9000 Ghent, Belgium\\
$^{5}$Lorentz Institute for Theoretical Physics, Leiden University, PO Box 9506, 2300 RA Leiden, the Netherlands\\
$^{6}$Department of Physics and Astronomy, University of Exeter, Stocker Road, Exeter EX4 4QL, UK\\
$^{7}$Dipartimento di Fisica ``Giuseppe Occhialini'', Universit\`a degli Studi di Milano Bicocca, Piazza della Scienza, 3 I-20126 Milano MI, Italy\\
$^{8}$Department of Astrophysics, University of Vienna, T\"urkenschanzstrasse 17, A-1180 Vienna, Austria\\
$^{9}$Centre for Data Science, Artificial Intelligence and Modelling, University of Hull, Cottingham Road, Hull, HU6 7RX, UK\\
$^{10}$E. A. Milne Centre for Astrophysics, University of Hull, Cottingham Road, Hull, HU6 7RX, UK\\
}

\date{Accepted XXX. Received YYY; in original form ZZZ}

\pubyear{\the\year{}}

\begin{document}
\label{firstpage}
\pagerange{\pageref{firstpage}--\pageref{lastpage}}
\maketitle

\begin{abstract}
\textit{JWST} surveys reveal a greater space density of high-redshift UV-bright galaxies than predicted by conventional galaxy formation models. We present results from a $L=100\,\cMpc$ cosmological simulation evolved to $z=5$ with a variation of the COLIBRE galaxy formation model that adopts a density-dependent stellar initial mass function (IMF), such that stellar populations formed from dense gas are born with a top-heavy IMF. Crucially, heavy element and dust yields, and supernova feedback energetics, are self-consistently adjusted to the changing IMF. We model UV/optical emission (including nebular emission) from galaxies and its attenuation by dust. By allowing a significant fraction of high-redshift star formation to proceed with a top-heavy IMF, the rest-frame far-UV luminosities of early galaxies are elevated by up to a factor of $\simeq4$ with respect to the fiducial COLIBRE L100m6 simulation, which assumes a universal Chabrier IMF. This enables the formation of galaxies with observed brightness up to $M_{\rm UV} \simeq -20$ at $z=15$ (c.f. $M_{\rm UV} \simeq -18.5$ in the fiducial simulation), illustrating the potential of star formation with a top-heavy IMF to alleviate tensions with \textit{JWST} data. Later, the boost in far-UV emission is partly offset by attenuation due to increased dust surface densities from i) additional dust grain ejection from core-collapse supernovae and ii) efficient grain growth promoted by more metal-rich interstellar gas. The simulation reproduces the $z=5$ galaxy stellar mass function and rest-frame optical luminosity function with comparable accuracy to the fiducial simulation, and both simulations exhibit UV continuum slopes that are consistent with \textit{JWST} observations.
\end{abstract}




\section{Introduction}
\label{sec:intro}
Deep observations with the \textit{James Webb Space Telescope} (\textit{JWST}) have challenged our understanding of the early universe. Within its first years of operation, \textit{JWST} has identified a surprisingly abundant population of rest-frame UV-bright ($M_{\rm UV} < -22$) galaxy candidates at $z \gtrsim 10$ \citep[e.g.][]{arrabal2023, finkelstein2023, casey2024, robertson2024, naidu2026}, pushing estimates of the redshift of cosmic dawn to $z \gtrsim 15$. Whilst some of these sources have since been identified as having a dominant contribution to their emission from a central black hole \citep[e.g.][]{maiolino2024}, and some $z>14$ candidate galaxies were revealed to be low-redshift interlopers with significant dust attenuation and strong emission lines \citep[e.g.][]{arrabal2023}, a number of compelling detections remain \citep[e.g.][]{carniani2024}. The existence of such UV-bright galaxies at early epochs is incompatible with most pre-\textit{JWST} predictions and, taken at face value, the most extreme cases have been interpreted as being in tension with a standard $\Lambda$CDM cosmogony when inferring their stellar masses under conventional assumptions for the initial mass function (IMF) and star formation history \citep[e.g.][]{casey2024,xiao2026}.

Recent efforts to model the rest-frame far-UV ($\lambda \simeq 1500\,$\AA) luminosity function (UVLF) of galaxies in the early universe with cosmological hydrodynamical simulations e.g. IllustrisTNG \citep{vogelsberger2020}; MilleniumTNG \citep{kannan2023}; FLARES \citep{vijayan2021}; SPHINX \citep{katz2023sphinx}; COLIBRE \citep{lu2026b}, and semi-analytic models e.g. GALFORM \citep{cowley2018, lu2025}; SHARK \citep{lagos2019}; DELPHI \citep{mauerhofer2023} and the Santa Cruz model \citep{yung2024, somerville2025}, broadly reproduce the observations at $z\leq 10$ (with varying degrees of accuracy), but, with the exception of \citet[][subsequently followed-up by \citealt{lu2025}]{cowley2018}, which
presented predictions prior to the launch of \textit{JWST}, these studies either do not present $z>10$ predictions for the UVLF, or reveal that the UV luminosity of galaxies at fixed space density is too low, particularly for the bright end of the UVLF. 

Various mechanisms have been advanced to explain the abundance of UV-bright galaxies at high redshift such as early star formation proceeding with a top-heavy stellar initial mass function \citep[IMF, first proposed by][see also \citealt{haslbauer2022, inayoshi2022, pacucci2022, cameron2024, woodrum2024, lu2025}]{cowley2018} or elevated star formation efficiency \citep{inayoshi2022, pacucci2022, dekel2023, mauerhofer2025, somerville2025}; bursty star formation \citep{mason2023, mirocha2023, shen2023, sun2023, kravtsov2024}; the efficient removal of obscuring dust by radiation-driven outflows \citep{ferrara2023, fiore2023, shen2023, ziparo2023, yung2024}; or a significant active galactic nucleus (AGN) contribution to galaxy luminosities \citep{pacucci2022, hegde2024}. In the absence of rest-frame optical diagnostics, it is difficult to discriminate between or rule out these possibilities with authority, but one can estimate the maximum `boost' to the observed far-UV luminosity that plausibly results from each physical mechanism \citep[e.g.][]{mason2023, wang2024}, and/or model their influence on the observable and physical properties of galaxies at early epochs \citep[e.g.][]{katz2025megatron} and in the evolved cosmos \citep[e.g.][]{barber1,barber2,barber3}. \citet{lu2025}, for example, show that invoking a top-heavy IMF in starbursts \citep[per the model of][]{cowley2018} roughly reproduces the observed UVLF up to $z=14$ when the model is modified to include a metallicity-dependent timescale for the formation of dust from metals in the interstellar medium.

The IMF describes the initial mass distribution of stars comprising a stellar population. It is of critical importance as it influences the energetics of stellar feedback, the metal and dust yields returned to the interstellar medium, and the population's spectral energy distribution. The most widely used functional forms of the IMF include the single power-law of \cite{salpeter}, and the more recent \cite{kroupa} and \cite{chabrier} forms that adopt a shallower slope or `rollover' in the low-mass regime ($m < 0.5\,\Msun$), as first proposed by \citet{Scalo1986}. These forms of the IMF are based upon observed resolved star counts within the Solar neighbourhood and are commonly referred to as `Milky Way-like' IMFs. In the absence of resolved star counts beyond the Local Group, the IMF is most frequently assumed to be universal. However, this is a strong assumption, and variation of the IMF is readily motivated. Observational evidence for systematic variations in the IMF on a range of spatial and mass scales pre-dates \textit{JWST} \citep[e.g.][]{rieke1993,massey1995,hillenbrand1997,larson1998}, and SED-fitting to spectroscopic data from \textit{JWST} indicates top-heavy IMFs at early times \citep[e.g.][]{bekki2023,cameron2024}. Variation of the IMF is also motivated by numerical simulations of molecular cloud evolution \citep[e.g.][]{hocuk2012,bate2025}. An early argument for IMF variations based on the properties of high-redshift galaxies was set out by \citet[][see also \citealt{lacey2016}]{baugh2005}, who found that their GALFORM semi-analytic galaxy formation model was only able to explain the observed numbers of faint sub-mm galaxies at $z \sim 1-4$ if the IMF in high-redshift starbursts was assumed to be top-heavy.

Particular interest in systematic variations in the IMF was sparked by a flurry of studies inferring central mass-to-light ($M/L$) ratios of massive elliptical galaxies that appear incompatible with Solar neighbourhood IMFs \citep[e.g.][see \citealt{smith2020} for a comprehensive review and critical appraisal of the evidence]{vandokkum2008, treu2010, vandokkum2010, cappellari2012, cappellari2013a, lu2024}. The presence of strong absorption features sensitive to low surface gravity stars in these systems \citep[e.g.][]{conroy2012,labarbera2013} favours a bottom-heavy IMF implying excess dwarf stars rather than excess remnants of massive stars \citep[see also][]{Cheng2026}. However, \citet{PvD2024} recently proposed an IMF that is both bottom- and top-heavy in high velocity dispersion regions, as a means to explain the high $M/L$ of present-day massive ellipticals and the low $M/L$ of their high-redshift progenitors. Indirect corroborative evidence from local galaxies also suggests that the IMF becomes top-heavy during periods of high star formation rate \citep[e.g.][]{meurer2009,gunawardhana2011}, as is also inferred from galaxy-wide IMF analyses \citep{kroupa2003,yan2017,jerabkova2018}.

From a theoretical perspective, a universal IMF across diverse star-forming environments is counter-intuitive, since gas cloud fragmentation during star formation depends on local temperature and density through the Jeans mass \citep{larson1998, bonnell2006, hennebelle2024}. Recent simulations corroborate this expectation \citep{chon2022, bate2023}. The temperature of a gas cloud for a given density depends on its metallicity \citep{omukai2005}, which induces a relation between metallicity and the IMF, such that metal-poor gas clouds are expected to form top-heavy stellar populations \citep{abel2002, bromm2002}, an effect of particular relevance for the formation of Population III stars \citep[e.g.][]{sharda2020}. Likewise, the redshift evolution of the cosmic microwave background radiation implies an evolving gas temperature floor, $T_{\mathrm{b}}\propto(1+z)$, that inhibits cloud fragmentation \citep[e.g][]{schneider2010, chon2022, bate2025}. For a detailed review of the physical drivers that can influence the shape of the IMF, see \citet{hennebelle2024}. 

Significant systematic variation of the IMF induces non-trivial consequences for the physical and observable properties of galaxies and, by extension, the interpretation of observable properties \citep[e.g.][]{baugh2005,fontanot2014,clauwens2016,lacey2016,guszejnov2017,barber1,gutcke2019,cueto2024,lu2025,fontanot2026}. Whilst increasing the relative number of massive stars within a stellar population will, for some period, increase the population's far-UV luminosity, it also increases pre-explosive feedback, i.e radiation pressure, momentum injection from stellar winds and thermal energy from photoheating within \HII\ regions. More significantly, it also increases the number of stars with mass $>8$ M$_\odot$, which are generally assumed\footnote{There remains some debate whether all such stars yield CCSNe \citep[e.g.][]{smartt2009}, or instead experience `dark deaths' such as direct collapse to a black hole \citep[e.g.][]{heger2010, sukhbold2016}.} to end their lives as core-collapse supernovae (CCSNe), injecting $\sim 10^{51}\,{\rm erg}$ into the interstellar medium (ISM) and thus more strongly regulating subsequent star formation. The additional CCSNe will also increase the yield of metals ejected into the ISM (increasing its rate of radiative cooling) and that of dust, absorbing a fraction of the additional UV photons, though some of the additional dust is expected to be destroyed in the SN reverse shock \citep{kirchschlager2019}. Varying the IMF also impacts metal and dust yields from intermediate mass ($\simeq 0.8-8\,\Msun$) asymptotic giant branch (AGB) stars \citep{ferrarotti2006}, and the number of ionising photons produced by young massive stars \citep{robertson2022}, with potentially significant consequences for the epoch of reionisation. 

Self-consistently modelling the diverse physical and observable effects of IMF variations is therefore vital for assessing whether they can alleviate the shortcomings of current galaxy formation models in the high-redshift regime, and do so without introducing fresh tensions with complementary observables. However, such self-consistency is not trivial to implement and is often overlooked. As noted above, \cite{lu2025} found that the GALFORM semi-analytic model, which assumes a top-heavy IMF in starbursts, yields good agreement to the observed UVLF at $z\gtrsim 10$ only if attenuation by dust at these early times is reduced or negligible due to a finite dust formation timescale. However, their IMF implementation only self-consistently adjusts stellar luminosities and metal yields without adjusting the energetics of outflows in response to the changing number of CCSNe per unit stellar mass formed. \cite{trinca2024} conclude that a metallicity and redshift-dependent IMF model improves the fit of the Cosmic Archaeology Tool (CAT) semi-analytic model to observed galaxy UV luminosity functions at $z>9$, but their model does not adjust stellar feedback, metal yields or ionising photon production to account for IMF changes. \cite{cueto2024} examined the influence on the high-redshift UVLF of imposing a top-heavy IMF on all stellar populations forming at early epochs, using a version of the ASTRAEUS semi-analytic model that self-consistently adjusts the energy injection and metal yields of CCSNe in response to IMF variation. They concluded that the stronger feedback produced by top-heavy populations offsets their enhanced UV luminosity per unit mass, resulting in little change to the UVLF. However, a follow-up study by \cite{hutter2025}, in which the ASTRAEUS model was modified so the top heaviness of the IMF is an increasing function of a galaxy's specific star formation rate (sSFR), found good agreement with the $z>10$ UVLF. \cite{fontanot2026} similarly implemented a self-consistent (in feedback and metal ejection) SFR density-dependent IMF in the GAEA semi-analytical model and find better agreement with the observed $z>10$ UVLF data than for a Milky Way-like IMF. 

The impact of self-consistent modelling of IMF variations on galaxy population statistics such as the high-redshift UVLF has yet to be explored with cosmological hydrodynamical simulations of large volume. It is of critical importance to do so, because such simulations rely on fewer approximations and assumptions than semi-analytic models. In particular, simulations with a live dust treatment are able to model the impact of non-homogeneous dust distributions on attenuation. Modelling the impact of IMF variations on the high-redshift galaxy population presents a particular challenge, as it places competing demands on computational power. Observed galaxy populations are naturally biased towards brighter sources by detection limits: at present the UVLF at $z\simeq 10$ is characterised only for galaxies with rest-frame far-UV magnitudes $M_{\rm UV} < -16$. Brighter galaxies at early epochs likely form in regions representing rare peaks of the density field \citep[e.g.][]{crain2009,lovell2021}, motivating the use of large simulation domains to improve the sampling of large-scale modes in the power spectrum. On the other hand, modelling the emergence of the nascent galaxy population necessitates the adoption of high resolution to capture the formation of the first stellar populations at relatively low local overdensity: current estimates of the stellar mass of $z\simeq 12$ galaxies are as low as $\sim10^8$ M$_\odot$ \citep{harvey2025}, placing an upper limit on the required baryonic mass resolution of simulations of at least an order of magnitude below this scale.

We therefore examine the influence of a density-dependent variable top-heavy IMF on the physical and observable properties of high-redshift galaxies, using new simulations evolved with a modified version of the galaxy formation model developed for the COLIBRE suite of simulations \citep{colibre1, colibre2}. COLIBRE directly models the multiphase interstellar medium, including the evolution of dust grains of different sizes and chemical compositions. The COLIBRE suite includes cosmological volumes of side $L=25-400\,\cMpc$ at particle mass $\sim 10^5-10^7\,\Msun$ (for both baryons and dark matter), and the parameters of the models governing feedback processes have been calibrated to reproduce at $z=0$ the galaxy stellar mass function, and the sizes and black hole masses of galaxies. The fiducial simulations, which adopt a universal \citet{chabrier} IMF, reproduce the mass and star formation rates of galaxies inferred from observations at $0\leq z \leq 10$ \citep{chaikin2025}, but yield fewer UV-bright galaxies than observed at $z \gtrsim 10$ \citep{lu2026b}, assuming reasonable redshift error in the data. This shortcoming motivates examination of a potential remedy that yields increased far-UV photon production at fixed stellar mass: we therefore develop a version of the COLIBRE model with a variable IMF that is driven by gas density, which correlates with pressure, SFR density and local radiation and cosmic ray intensities. The variable IMF takes the Solar neighbourhood form for stellar populations forming from moderate-density gas and becomes increasingly top-heavy for higher-density natal gas. The energetics of stellar feedback, and the yields of heavy elements and dust, are self-consistently adjusted in response to changes of the IMF. Our aim is not to `calibrate' the IMF to reproduce the high-redshift UVLF, but rather to examine whether a variable IMF can plausibly remedy the UV shortfall of $z \gtrsim 10$ galaxies without compromising the broad agreement of the fiducial COLIBRE simulations with many other observables at later epochs. 

The remainder of this paper is structured as follows. Section \ref{sec:methods} describes the simulation framework, our variable IMF model and methods for modelling galaxy spectra for our simulated galaxy populations. Section \ref{sec:validation} presents validation tests of the variable IMF simulation at $z=5$. Section \ref{sec:results} presents comparisons at $z\geq5$ between the variable IMF simulation and its fiducial COLIBRE counterpart. We summarise our findings in Section \ref{sec:summary}.

\section{Methods}
\label{sec:methods}
\subsection{The COLIBRE simulations}
\label{subsec:methods_colibre}

We introduce new simulations evolved with a version of the COLIBRE galaxy formation model that has been modified to incorporate a variable IMF rather than the fixed \citet{chabrier} IMF adopted by the fiducial model. Throughout this work, we compare results from the new simulations with their fiducial COLIBRE counterpart, starting from the same initial conditions. Comprehensive details of the COLIBRE model and its fiducial simulations are provided by \cite{colibre1}, and a thorough explanation of how the model's parameters were calibrated to reproduce select properties of the local galaxy population are provided by \citet{colibre2}. We briefly summarise below the elements of the fiducial model that are most relevant to our study, and then describe the variable IMF implementation in \S\ref{subsec:methods_vimf_model}, as well as the simulations used in this work in \S\ref{subsec:methods_sims}. 

Initial conditions for the simulations were generated with the \textsc{monofonIC} software \citep{monofonic1,monofonic2} at $z = 63$, using second-order Lagrangian perturbation theory. The simulations assume a $\Lambda$CDM cosmogony with parameters based on the maximum posterior likelihood values of the ‘3×2pt + all external constraints’ model derived from year three Dark Energy Survey data \citep[DES Y3,][]{abbott2022}: $\Omega_{\rm M,0} = 0.306$, $\Omega_{\rm b,0} = 0.0486$, $\sigma_8 = 0.807$, $h = 0.681$, $n_{\rm s} = 0.967$, with a single massive neutrino species of mass $0.06\,\eV$. The initial conditions comprise gas and dark matter particles of comparable mass (rather than comparable number) to minimise spurious energy transfer from dark matter to the baryons \citep{ludlow2021}. COLIBRE uses ‘partially fixed initial conditions’ \citep{angulo2016} whereby the amplitudes of modes with $(kL)^2 < 1025$ are fixed to the mean variance, where $k$ is the wavenumber and $L$ is the comoving box side length. The COLIBRE suite currently comprises simulations of three resolutions termed m5, m6 and m7, with initial baryon particle mass, $m_{\rm g}$, of $2.3 \times 10^5\,\Msun$, $1.8 \times 10^6\,\Msun$, and $1.5 \times 10^7\,\Msun$, respectively. The corresponding DM particle masses, $m_{\rm DM}$, are $3.0\times 10^5\,\Msun$, $2.4 \times 10^6\,\Msun$, and $1.9 \times 10^7\,\Msun$. The simulations at m5, m6 and m7 resolution have been evolved to $z=0$ in cubic volumes with a side length up to $L=25\,\cMpc$, $L=200\,\cMpc$ and $L=400\,\cMpc$, respectively.
 
The simulations were evolved using a version of \textsc{swift} \citep{swift} that combines the open source 2025.04 release of the software, using the SPHENIX smoothed-particle hydrodynamics scheme \citep{sphenix}, with new modules for subgrid treatments of unresolved physical processes. COLIBRE introduces several key improvements to hydrodynamical galaxy formation simulations, including a multiphase interstellar medium without a temperature floor, allowing gas to cool to $T\sim 10\,{\rm K}$; a live dust model \citep{trayford2025} that we discuss further in \S\ref{subsec:methods_dust}; and non-equilibrium gas cooling \citep{ploeckinger2025,richings2014a,richings2014b}. A gravitational instability criterion is used to identify gas particles eligible for star formation, at a rate governed by the \cite{schmidt1959} law, assuming a fixed star formation efficiency per free-fall time of $\epsilon$ = 0.01 \citep[see][]{nobels2024}.

Stellar particles represent simple stellar populations (SSPs) of the \citet{chabrier} IMF with lower and upper stellar mass limits of $0.1\,\Msun$ and $100\,\Msun$ respectively. Stellar particles enrich surrounding gas with metals via six channels: AGB stars, type-Ia SNe, CCSNe, neutron star mergers, common envelope jet SNe, and collapsars. We assume that stars more massive than $40\,\Msun$ do not enrich their surroundings \citep{correa2026}. Early non-explosive stellar feedback (radiation pressure, momentum injection from stellar winds and thermal energy from photoheating within \HII\ regions) is implemented as per \cite{benitez2025}. Energy feedback from CCSNe is implemented via stochastic thermal+kinetic injection, based on a version of the \cite{chaikin2023} model with modifications detailed by \citet[][see their \S3.7]{colibre1}. Supermassive black holes (SMBHs) are seeded in dark matter haloes (identified with an on-the-fly friends-of-friends algorithm) when they exceed a mass of $5\times 10^{10}\,\Msun$ at m7 resolution or $10^{10}\,\Msun$ at m6 and m5 resolutions, by converting the halo's densest gas particle into a BH particle. SMBHs accrete at a rate specified by the Bondi-Hoyle-Lyttleton formula, modified to account for turbulence and vorticity corrections \citep{krumholz2006}. BH-BH mergers are modelled per the method of \citet{bahe2022}. Thermal AGN feedback is implemented deterministically \citep[see e.g.][]{booth2009}, by heating the gas-neighbouring BH particles by a temperature increment proportional to the subgrid BH mass \citep{colibre1}.

The parameters governing feedback in COLIBRE were calibrated to ensure reproduction of the $z=0$ galaxy stellar mass function, galaxy sizes, and galaxy SMBH masses \citep{colibre2}. Of particular relevance for this study, the strength of CCSN feedback is an increasing, monotonic function of natal (thermal) gas pressure, $f_{\rm E}(P_{\rm birth})$, where $f_{\rm E}$ is the energy of a single SN that couples to the ISM, in units of $10^{51}\,{\rm erg}$. Calibration against these $z=0$ observables enables the simulations to reproduce many other properties of the galaxy population, including the inferred mass and star formation rates of galaxies as early as $z\simeq 12$ \citep{chaikin2025}. The simulations are therefore an excellent foundation from which to examine the effects of variable IMFs on the formation and evolution of the galaxy population over cosmic history. However, we note that the power of the simulations to yield accurate predictions of the properties of the galaxy population at early epochs is hampered by the use of a simple flash reionisation model \citep{ploeckinger2025}, and the absence of an explicit model of the formation and evolution of Population III stars. We anticipate that these effects are primarily significant for the properties of very faint galaxies, which are not the focus of our study. 

Dark matter haloes are identified on-the-fly using a FoF algorithm, and subhaloes are identified in post-processing using \textsc{HBT-HERONS} \citep{herons}. A diverse range of galaxy properties are computed by aggregating the properties of particles comprising galaxies, using the Spherical Overdensity and Aperture Processor \citep[SOAP;][]{soap}. This includes dust-free galaxy luminosities, which we calculate within a projected circular aperture of radius $50\,\pkpc$ along a single random viewing angle. See \S\ref{subsec:methods_fsps} for methods of modelling emission from stellar particfles. We use a projected aperture for consistency with the dust-attenuated luminosities that we obtain from SKIRT (see \S\ref{subsec:methods_dust}). We use a radius of $50\,\pkpc$ for consistency with the methods introduced by \citet{colibre1}, but note that the sizes of high-redshift galaxies are much smaller than this size, so in practice we consider all gravitationally-bound stellar particles.

\subsection{Variable IMF implementation}
\label{subsec:methods_vimf_model}

We adopt an approach to varying the IMF similar to that of \citet{barber1}, who modified the EAGLE model so that the IMF deviated monotonically from the Solar neighbourhood form with increasing natal gas pressure. Those authors examined both top- and bottom-heavy IMFs, whilst we restrict ourselves to the former. We share their philosophy that the IMF should converge on the Solar neighbourhood form for star formation conditions similar to those of the Milky Way, and that variation of the IMF should be driven by a physically-meaningful property such as local gas conditions, as opposed to global properties that cannot directly influence the birth clouds of stellar populations. 

The fiducial COLIBRE simulations adopt the \citet{chabrier} IMF, but we follow \citet{barber1} in adopting the \citet{kroupa} IMF as the Solar neighbourhood form, as it is effectively indistinguishable from the former over the mass range $0.1-100\,\Msun$ and, being simply defined by a piecewise double power law, it is easier to use: $\md n/\md M \propto M^{x}$, where $x = -1.3$ and $-2.3$ for stellar masses below and above $0.5\,\Msun$, respectively. There are several easily implemented ways of increasing the number of massive stars relative to this form of the IMF, such as increasing the IMF upper mass cut-off \citep[e.g.][]{ma2025}, adjusting the pivot mass separating the low- and high-mass regimes \citep[e.g.][]{fontanot2018,cueto2024,woodrum2024,fontanot2026} or, as we adopt here, varying the slope of the IMF in the high-mass regime \citep[e.g.][]{barber1, cueto2024, lu2025}. \citet{barber1} found that, for top-heavy IMFs, adjusting the high-mass slope yields similar results to adjusting the pivot mass. We do not explore other methods of varying the IMF here for simplicity and note that our aim is not to constrain the shape of the IMF in the early universe, rather it is to assess the implications of increasing the number of massive stars at this epoch.

We therefore fix i) the lower and upper mass limits of the IMF to $0.1\,\Msun$ and $100\,\Msun$ respectively; ii) the pivot mass to $0.5\,\Msun$; and iii) the gradient of the IMF in the low-mass regime to $-1.3$. We only vary the gradient of the IMF in the high-mass regime, $\alpha$, allowing it to increase from a minimum value of $-2.3$ as a monotonic function of the density of natal gas, $n_{\rm H,birth}$:
\begin{equation}
    \frac{{\rm d}n}{{\rm d}M} \propto \left\{
    \begin{array}{ll}
        M^{-1.3} & {\rm where}\ \ 0.1 \leq M/\Msun < 0.5 \\
        M^{\alpha(n_{\rm H,birth} )} & {\rm where}\ \ 0.5 \leq M/\Msun < 100 .
    \end{array}
    \right.
    \label{eqn:slope_piecewise}
\end{equation}
We choose to vary $\alpha$ as a function of natal density in part because a relationship between the IMF and star formation conditions is supported by observations of local star-forming regions \citep[][]{meurer2009,gunawardhana2011,watts2018,zhang2018}. The choice of density is heuristic, and it yields more top-heavy star formation at earlier times because the average matter density is higher. We choose to use natal density, which spans $\sim7$ orders of magnitude, rather than the natal pressure because in lower resolution COLIBRE simulations ($m_{\rm g} \gtrsim 10^6 \rm \,\Msun$), even warm gas ($T \sim 10^4\,\K$) can be Jeans unstable and hence eligible to turn into stellar particles. The transition between cold and warm phases is locally sharp at this resolution, enabling neighbouring gas particles with similar gas densities to exhibit pressures that differ by several orders of magnitude.

Following \citet{barber1}, we vary $\alpha$ using a simple, monotonic sigmoid function:
\begin{equation}
    \alpha(n_{\rm H,birth}) = \alpha_{\rm{high}} + \frac{\alpha_{\rm{low}} - \alpha_{\rm{high}}}{ 1+ \left(n_{\rm H,birth}/n_{\rm H,pivot} \right)^\gamma}
    \label{eqn:sigmoid}
\end{equation}
where $\alpha_{\rm{low}}=-2.3$ and $\alpha_{\rm{high}}=-1.6$ are the extreme values of the gradient of the IMF at low and high birth densities, respectively. The parameters $\gamma$ and $n_{\rm H,pivot}$ determine the width of the sigmoid function and its pivot density, respectively. The choice of $\alpha_{\rm{low}}=-2.3$ ensures the IMF converges to the \citet{kroupa} form for $n_{\rm H,birth} \ll n_{\rm H,pivot}$. The choice of $\alpha_{\rm{high}}=-1.6$, and the adopted values of $\gamma$ and $n_{\rm H,pivot}$ are motivated in \S\ref{subsec:methods_imf_parameters}.

\begin{figure}
    \centering
    \includegraphics[width=0.9\columnwidth]{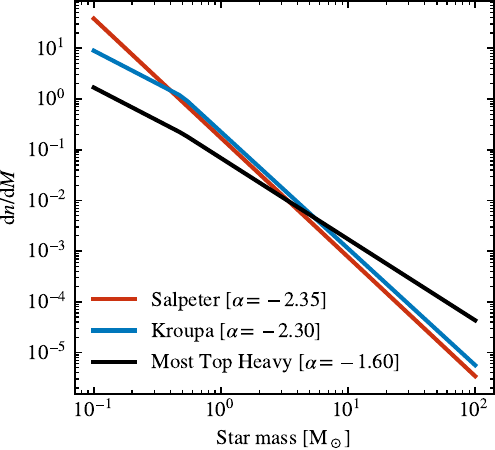}
    \caption{Common functional forms of the stellar initial mass function (IMF) shown for the mass range $0.1 < m_{\star} < 100\,\Msun$ and normalised such that each IMF integrates to $1\,\Msun$: those of \protect\citet[][red]{salpeter} and \protect\citet[][blue]{kroupa}; note that the \protect\cite{chabrier} IMF is not shown because it is indistinguishable over this mass range from that of \protect\cite{kroupa}. The black curve corresponds to the most top-heavy IMF used in this study ($\alpha = -1.6$, Eq. \ref{eqn:slope_piecewise}).}
    \label{fig:imf}
\end{figure}

Common forms of the IMF derived from resolved stellar number counts in the Solar neighbourhood are shown in Fig. \ref{fig:imf}: \citet[][red curve]{salpeter} and \citet[][blue curve]{kroupa}. The \cite{chabrier} IMF is not shown because it indistinguishable from the \cite{kroupa} IMF on this scale. The black curve shows the most top-heavy IMF we adopt, with $\alpha = \alpha_{\rm high} \equiv -1.6$. In each case, the IMF is normalised so that the stellar mass integrates to $1\,\Msun$ between the limits $0.1-100\,\Msun$. COLIBRE's subgrid implementations of energy feedback from CCSNe, heavy element enrichment and dust formation each begin from a parameterised form of the IMF, meaning that self-consistent evolution of these aspects of the model follow without the need for further modification of the code. We note however that contributions from early stellar feedback, as detailed by \cite{benitez2025}, are not self-consistently adjusted and assume a Chabrier IMF at all times. The energy injection rates in the early stellar feedback model are pre-tabulated, and self-consistently adjusting them as the IMF varies would require a significant change to the code.

\subsubsection{Variable IMF parameters and modification of CCSN
feedback}
\label{subsec:methods_imf_parameters}

The fiducial COLIBRE model, adopting a universal \citet{chabrier} IMF, injects into the ISM an energy per CCSN that is a monotonically-increasing function of the thermal pressure of natal gas, $f_{\rm{E}}(P_{\rm{birth}})$, so that stellar populations born from gas with greater thermal pressure inject more energy per CCSN. This energy is specified by a sigmoid function whose parameters were calibrated to reproduce the $z=0$ properties of the galaxy population \citep[see][]{colibre2}. The energy, in units of $10^{51}\,\erg$, asymptotes to $0.3$ at low pressure (for m6 resolution) and $4.0$ at high pressure. \citet{colibre1} remark that this has multiple interpretations, including that it is an ingredient necessary to overcome residual numerical overcooling, or on physical grounds, because some CCSNe inject kinetic energies greater than $10^{51}\,\erg$ \citep{mazzali2014} or because the IMF may become top-heavy in high pressure star forming regions \citep[e.g.][]{meurer2009,gunawardhana2011}. The latter is appealing in light of the independent arguments for a top-heavy IMF in high-redshift galaxies, but the fiducial COLIBRE simulations do not model the other important consequences of the IMF becoming top-heavy, namely the stellar luminosities and (time-dependent) change of the metal and dust yields. We are therefore motivated to examine whether a realistic galaxy population can be obtained in a simulation that uses a fixed $f_{\rm E}$ (energy injected per CCSN) but achieves the same effect as COLIBRE's empirical $f_{\rm E}(P_{\rm birth})$ relation by varying the number of CCSNe per unit stellar mass formed via variations in the IMF, rather than varying the energy injected per CCSN.

\begin{figure}
    \centering
    \includegraphics[width=\columnwidth]{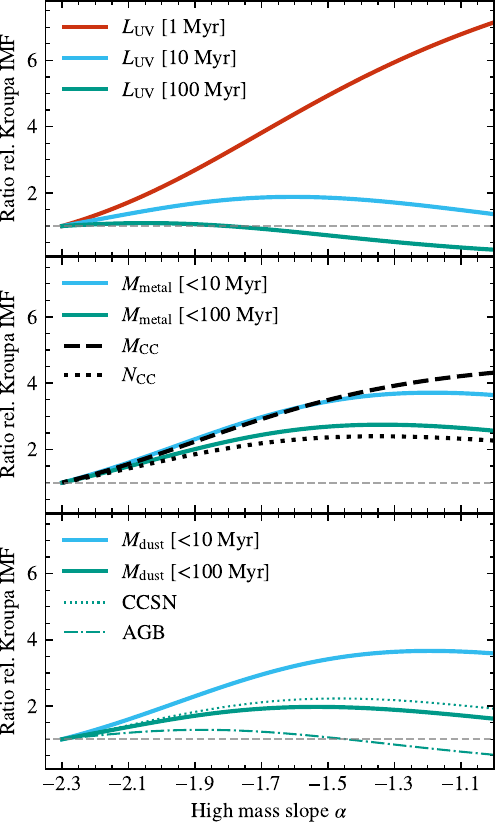}
    \caption{\textit{Top:} The far-UV luminosity of a simple stellar population (SSP) with solar metallicity ($Z=0.0134$) at ages $(1,10,100)\,\Myr$ (red, cyan and green curves, respectively), as a function of the high-mass slope of the IMF ($\alpha$, parameterised as Eq. \ref{eqn:slope_piecewise}), normalised by the far-UV luminosity of a coeval SSP with a \citet{kroupa} IMF. \textit{Middle:} the mass fraction (black dashed curve) and number (black dotted curve) of the population contributed by stars of mass $8 < m_\star < 100\,\Msun$, which are assumed to be CCSNe progenitors, as well as the cumulative mass of metals ejected by the same SSPs as in the top panel at ages $(10,100)\,\Myr$ (solid cyan and green curves). \textit{Bottom:} cumulative mass of dust ejected at the same ages, with the contributions to the 100 Myr total from CCSNe and AGB stars denoted by dotted and dot-dashed curves, respectively. The horizontal dashed line on each panel shows where the SSP property is equal to that of a \citet{kroupa} IMF.}
    \label{fig:imf_boosts}
\end{figure}

Stellar populations with a top-heavy IMF ($-1.8 \lesssim \alpha \lesssim -1$) yield roughly twice as many CCSNe as those with a Kroupa IMF, assuming that all stars of mass greater than $8\,\Msun$ are the progenitors of CCSNe. This is demonstrated by the black dotted curve in Fig. \ref{fig:imf_boosts} (middle panel), which shows how the number of CCSNe varies with $\alpha$, relative to the number of CCSNe assuming a Kroupa IMF. We therefore adopt a fixed energy per CCSN of $2 \times 10^{51}$ erg (i.e. $f_{\rm{E}}=2$), so that top-heavy stellar populations that form from gas of high density inject broadly the same total CCSNe energy per unit stellar mass formed, $\simeq 5\times 10^{49}\,\erg\,\Msun^{-1}$, as is the case in the fiducial COLIBRE model with a CCSN energy of $10^{51}\,\erg$ and ${\rm max}[f_{\rm E}(P_{\rm birth})] = 4$. As noted in \S\ref{sec:intro}, we do not seek to calibrate the form of the IMF, our aim is to examine whether an IMF that becomes top-heavy in high-redshift galaxies can help to alleviate the tension between the observed UVLF and current galaxy formation models, without inducing fresh tensions with complementary observables, either in the high-redshift cosmos or at later times. We adopt a maximum gradient of the high-mass slope of the IMF of $\alpha_{\rm{high}} = -1.6$, on the basis of balancing a boost in the UV luminosity of very young ($\lesssim1$ Myr) stellar populations with a decline in UV luminosity after $\simeq100$ Myr resulting from the death of massive stars. The top panel of Fig. \ref{fig:imf_boosts} shows that $\alpha = -1.6$ yields stellar populations that are a factor of ($\simeq 4.4, 1.9)$ as UV-bright as those formed with a Kroupa IMF for population ages of $(1,10)\,\Myr$ at solar metallicity, while an even shallower slope of $\alpha>-1.6$ reduces both the $10$ and $100$ Myr instantaneous far-UV luminosity. Fig. \ref{fig:imf_boosts} and the time evolution of the far-UV luminosity are discussed further in \S\ref{subsec:methods_fsps}. 

The remaining parameters in Eq. \ref{eqn:sigmoid} were chosen to be $\gamma=0.65$ and $n_{\rm H,pivot}=30\,\percc$, based on analyses of several $L=50\,\cMpc$ simulations at m6 resolution, evolved to $z=5$, that span plausible ranges of these parameters. A primary aim of the variable IMF model was to yield a redshift evolution of the CCSNe energy injected per stellar mass formed that is comparable to that of the fiducial COLIBRE model, a corollary of which is that the variable IMF simulation should reproduce the $z=5$ galaxy stellar mass function (GSMF) of the fiducial simulation. Had this not proven the case (which we show in \S\ref{sec:validation}), it would have been necessary to adjust the variable IMF model. 

\subsection{Variable IMF simulations}
\label{subsec:methods_sims}

\begin{table*}
    \centering
    \caption{Key parameters of the fiducial COLIBRE and variable IMF L100m6 simulations. Columns are as follows: the simulation label; the comoving box side length, $L$; the initial mean baryonic particle mass, $m_{\rm g}$; the initial mean CDM particle mass, $m_{\rm CDM}$; the initial number of baryonic particles, $N_{\rm b}$; the initial number of CDM particles, $N_{\rm CDM}$; the comoving gravitational softening length, $\epsilon_{\rm com}$; the adopted IMF; the adopted feedback energy injected per CCSN. The proper gravitational softening length is capped at a maximum of $0.7$ pkpc which applies for $z<1.57$.}
    \begin{tabular}{lrrrrrrlll} 
        \hline
        Label & $L_{\rm{box}}$ & $m_{\rm{g}}$  & $m_{\rm{CDM}}$ & $N_{\rm{b}}$ & $N_{\rm{CDM}}$& $\epsilon_{\rm{com}}$ & IMF & Energy per CCSN, $f_{\rm{E}}$ \\
         & [cMpc] & [$10^6\,\Msun$] & [$10^6\,\Msun$] &  & & [ckpc] &  & [$10^{51}\,{\rm erg}$]\\
        \hline
        Fiducial & $100$ & $1.84$ & $2.42$ & $1504^3$ & $4\times1504^3$ & 1.8 & \citet{chabrier} & Variable \citep[Eq. 2 of][]{colibre2}  \\
        Variable IMF & 100 & $1.84$ & $2.42$ & $1504^3$ & $4\times1504^3$ & 1.8 & Variable (eqs. \ref{eqn:slope_piecewise} \& \ref{eqn:sigmoid}) & $2$ \\
        \hline
    \end{tabular}
    \label{tab:sims}
\end{table*}

We have conducted a simulation using the variable IMF implementation that uses the same $L=100\,\cMpc$ initial conditions at m6 resolution as the fiducial L100m6 COLIBRE simulation, evolved to $z=5$. We stop the simulation at this epoch primarily to limit its computational expense and because it is possible to compare the simulations with a GSMF inferred from rest-frame optical observations at $z=5$. We record the state of the simulation at 42 redshifts between $z = 20$ and $z=5$, saving 16 full snapshots at all integer redshifts and 26 `snipshots' containing a reduced set of output data. This differs slightly from that of the fiducial simulation, which saves 45 outputs over the same redshift range with 12 full snapshots and which does not include all integer redshifts for $z>10$. Numerical parameters of the simulation, besides those discussed in \S\ref{subsec:methods_vimf_model}, are identical to those adopted by the fiducial m6 simulations. Comparison of the variable IMF simulation with its counterpart from the fiducial COLIBRE suite therefore enables the influence of the IMF on the simulated galaxy population to be isolated. The influence of simulation box size on the UV luminosity function at $z=15$ is shown in Appendix \ref{appendix:convergence}. Key details of the fiducial COLIBRE and variable IMF L100m6 simulations are summarised in Table \ref{tab:sims}. The evolution of the characteristic high-mass IMF slope of newborn stellar populations that emerges from the L100m6 variable IMF simulation is shown in Appendix \ref{appendix:alpha_evolution}.

As noted in \S\ref{subsec:methods_colibre}, the fiducial COLIBRE simulations use \textsc{HBT-HERONS} \citep{herons}, an updated version of \textsc{HBT+} \citep{hbt,hbtplus}, to identify subhaloes in post-processing. As the variable IMF simulation was conducted prior to the release of HBT-HERONS, we use HBT+ for the variable IMF simulation and HBT-HERONS for the fiducial simulation. Distinctions between the two halo finders are in the treatment of substructure and have a negligible impact on results presented here. As both \textsc{HBT+} and \textsc{HBT-HERONS} track subhaloes in time rather than using only instantaneous phase-space information, their results are sensitive to the temporal spacing between snapshots. However, the output times of the variable IMF and fiducial simulations are sufficiently similar to not introduce significant differences. As with the fiducial simulation, galaxy properties are computed using the Spherical Overdensity and Aperture Processor \citep[SOAP;][]{soap}.

\subsection{Emission properties of stellar populations}
\label{subsec:methods_fsps}

We treat stellar particles as SSPs, and obtain their spectral energy distributions (SEDs) and broadband luminosities on a particle-by-particle basis using the Flexible Stellar Population Synthesis \citep[FSPS][]{fsps} software with the Basel spectral library \citep{basel1,basel2,basel3} and Padova isochrones \citep{padova1,padova2}. For consistency with the implemented IMFs, we integrate over the stellar mass range $0.1-100\,\Msun$ rather than the FSPS default of $0.1-120\,\Msun$. Note that, while FSPS can include the effects of binary stars, this functionality is limited to use with a \citet{salpeter} IMF, so we do not account for emission from binaries. To obtain far-UV magnitudes, we assume a top-hat transmission function spanning $1450-1550$ \AA, broadly aligning with the width of JWST/NIRspec optical/near-IR filters used to characterise rest-frame far-UV luminosities of distant galaxies \citep[as used by e.g.][]{finkelstein2024}. We ignore AGN emission and account for nebular continuum and line emission from ionised gas in \HII\ regions using a fixed gas ionisation parameter of $\log_{10}U=-2$ (see \S\ref{subsec:methods_nebular_emission}). Although not relevant to our galaxy luminosity calculations, we note that the mass evolution of SSPs computed by FSPS can deviate from that of stellar particles in COLIBRE due to small differences in the adopted stellar lifetimes and stellar remnant masses. The difference increases with the age of the stellar population, and yields a median discrepancy in the mass of galaxies of mass-weighted stellar age $\sim 1\,\Gyr$ at $z=5$ of 0.07 (0.24) dex for the fiducial (variable IMF) simulation, with the greater discrepancy in the variable IMF simulation stemming from the greater mass fraction contributed by remnants.

To illustrate the boost of the intrinsic luminosity (i.e. that unattenuated by dust) that follows from making the IMF more top-heavy, the top panel of Fig. \ref{fig:imf_boosts} shows, as a function of the IMF slope in the high-mass regime, the instantaneous far-UV luminosity of a solar metallicity SSP at $(1,10,100)\,\Myr$ (red, cyan and green curves, respectively). The curves are normalised by the luminosities of a coeval SSP formed with the \citet{kroupa} IMF. We assume a solar metallicity of $\mathrm{Z}_\odot=0.0134$ \citep{asplund2009}, and note that solar metallicity is selected purely as an example and the choice of metallicity has little impact on the relative differences between the two IMFs. The middle and bottom panels illustrate the corresponding boosts to some of the quantities that can act to counteract the intrinsic far-UV luminosity boost: the number of stars of mass greater than $8\,\Msun$ (which in COLIBRE are assumed to be the progenitors of CCSNe), $N_{\rm CC}$ (dotted black curve); the corresponding initial mass in stars that are CCSNe progenitors, $M_{\rm CC}$ (dashed black curve); the total metal mass (middle panel) and dust mass (bottom panel) ejected for SSP ages of $(10,100)\,\Myr$ (cyan and green curves, respectively). Green dotted and dot-dashed curves show the contributions to the total mass of dust produced by CCSNe and AGB stars, respectively, at $100\,\Myr$ (at $10\,\Myr$ the dust ejection is exclusively from CCSNe). The curves do not account for the destruction of dust in CCSNe reverse shocks (which self-consistently adjusts in our simulations in response to variation of the IMF), nor do they account for dust growth in the ISM after ejection. We note that dust yields from AGB and CCSNe vary significantly in the literature, as demonstrated by \cite{trayford2025}. If we instead used the dust yields presented by \cite{dwek1998}, the dust-to-metal mass ratios from CCSNe could increase by over an order of magnitude, though subsequent dust grain growth in the ISM would likely dominate over this effect \citep{bakx2025}. 

Fig. \ref{fig:imf_boosts} illustrates that the far-UV luminosity boost with respect to the \citet{kroupa} IMF is briefly very high for young, top-heavy stellar populations (a factor of $7.1$ at $1\,\Myr$ for $\alpha=-1$) but, owing to the short lifetimes of high-mass stars, the effect is short-lived, and for ages $\gtrsim 10\,\Myr$ the boost ceases to be a monotonic function of $\alpha$. Note that such an increase in high-energy photons should impact early stellar feedback, but this is not self-consistently adjusted as the IMF varies. The maximum intrinsic far-UV luminosity boost (relative to a Kroupa IMF) for a $10\,\Myr$ old, solar metallicity population is $1.9$ for $\alpha \simeq -1.6$. Such a stellar population will also produce roughly twice as many CCSNe (assuming progenitor masses $8 < m_\star/\Msun < 100$), contributing to more effective regulation of subsequent generations of star formation, and will eject a factor of $3.2$ more mass of both metals and dust. Absorption of far-UV photons by the latter might compensate (or over-compensate) the UV luminosity boost resulting from the greater number of high-mass stars. As noted in \S \ref{subsec:methods_colibre} we assume that stars more massive than $40\,\Msun$ do not enrich the ISM. The boost to metal and dust masses from a top-heavy IMF would be greater still if these massive stars contributed to enrichment, but the expected yields from stars of these masses are uncertain.

\begin{figure}
    \includegraphics[width=\columnwidth]{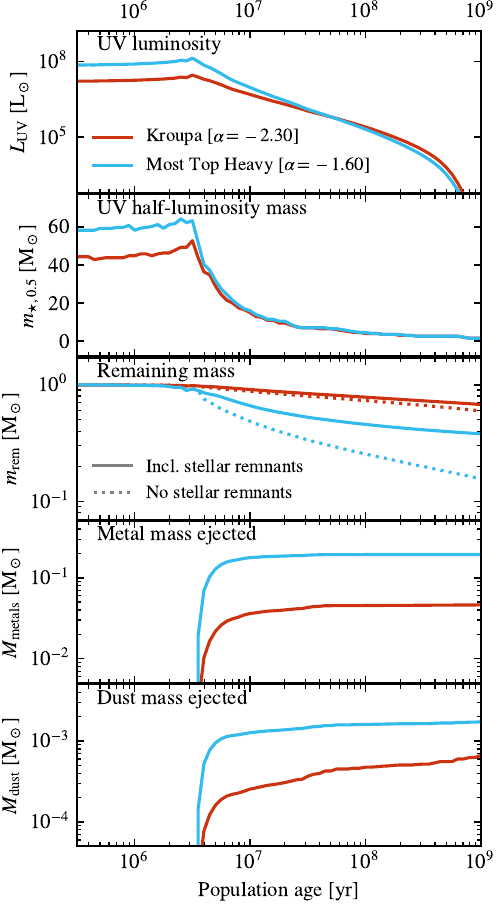}
    \caption{Evolution of the following properties of SSPs with Kroupa IMF (red curves) and the most top-heavy IMF used in our model (cyan curves), of solar metallicity ($Z=0.0134$) and initial mass $1\,\Msun$: i) the far-UV luminosity, ii) the UV half-luminosity mass (the mass of the star above, or below, which half of the far-UV luminosity is produced), iii) the mass of the SSP remaining in stars and remnants (solid curves), and stars only (dotted curves); iv) the cumulative metal mass ejected and v) the cumulative dust mass ejected.}
    \label{fig:time_evolved_ssp}
\end{figure}

We illustrate the time-dependence of consequences for SSPs stemming from changing the high-mass slope of the IMF in Fig. \ref{fig:time_evolved_ssp}. We show the evolution after a stellar population forms of i) the far-UV luminosity; ii) the star mass above which half of the total far-UV luminosity is produced (`UV half-luminosity mass'); iii) the mass of the SSP remaining in stars and remnants (solid curves), and stars only (dotted curves); iv) the cumulative metal mass ejected and v) the cumulative dust mass ejected. The SSPs have initial mass $1\,\Msun$ and solar metallicity. Results are shown for SSPs with a Kroupa (red curves) IMF and top-heavy IMF with a high-mass slope of $\alpha=-1.6$ (cyan curves). As in Fig. \ref{fig:imf_boosts}, the total dust mass ejected shown here does not account for dust destruction by CCSN reverse shocks, or dust growth in the ISM. Small features in the UV half-luminosity mass curves are a consequence of interpolating FSPS outputs with a finite time and mass resolution.

As previously inferred from Fig. \ref{fig:imf_boosts}, a top-heavy IMF with $\alpha=-1.6$ yields an initial increase in UV luminosity relative to a Kroupa IMF by a factor of up to $4.7$ for ages $\lesssim 20\,\Myr$. The brief increase in far-UV luminosity at an age of $\simeq 3\,\Myr$ is a consequence of massive stars experiencing a Wolf-Rayet phase. For ages $\gtrsim 40\,\Myr$, once the additional massive stars responsible for the top-heavy IMF's greater far-UV luminosity have died, the relative paucity of intermediate-mass stars in the top-heavy IMF results in its far-UV luminosity being slightly less than that of the Kroupa IMF. The relation between the IMF of young stars and the rest-frame far-UV luminosity of galaxies is therefore particularly sensitive to the star formation history. The UV half-luminosity mass, $m_{\star,{\rm 0.5}}$, is the initial star mass for which the far-UV luminosity produced by (remaining) stars in the interval [$0.1\,{\rm M}_\odot,m_{\star,{\rm 0.5}}$] is equal to that produced by those in the interval [$m_{\star,{\rm 0.5}},100\,\Msun$], which by definition is half of the population's total far-UV luminosity. For populations of age $\lesssim 3\,\Myr$, $m_{\star,{\rm 0.5}} \simeq 45\,\Msun$ for a Kroupa IMF and $\simeq 60\,\Msun$ for a top-heavy IMF with $\alpha=-1.6$, but $m_{\star,{\rm 0.5}}$ effectively converges for the two IMFs by $\simeq 6\,\Myr$, corresponding to the lifetimes of stars with mass $m_\star \simeq 33\,\Msun$. As the population ages and stars of progressively lower mass reach the end of their main sequence lifetime, the UV half-luminosity mass declines such that the far-UV luminosity is dominated by stars of ever lower mass: for both IMFs, $m_{\star,{\rm 0.5}} \simeq (15, 4)$ for population ages of $(10,100)\,\Myr$.

The bottom three panels of Fig. \ref{fig:time_evolved_ssp} illustrate the effect of mass loss from the stellar population as the constituent stars reach the end of their lifetimes and eject both metals (fourth panel) and dust (fifth panel); note that in our simulations metals may form dust grains in the ISM after ejection, depending on the elements released, but this is not accounted for in these figures. At a population age of $1\,\Gyr$, the stellar population retains a fraction of 0.68 (0.38) of its original mass for a Kroupa (top-heavy) IMF, with a fraction of 0.08 (0.22) of the initial mass being in the form of stellar remnants. As seen in Fig. \ref{fig:imf_boosts}, increasing the relative number of massive stars in the population results in the ejection of a greater mass of metals and dust into the ISM, which will also boost winds and radiation pressure. The ejected metal mass is mostly synthesised by CCSNe, with sub-dominant contributions from thermonuclear SNe and AGB stars, resulting in almost all of the metal mass synthesised by the population being ejected prior to the population reaching an age of $10\,\Myr$, for both IMFs. The increase in cumulative metal mass ejected by the top-heavy IMF relative to a Kroupa IMF is a factor of ($3.2$, $2.6$, $2.5$) at ages ($10$, $100$, $1000$) Myr, and the corresponding increase in cumulative dust mass ejected is a factor of ($3.2$, $2.0$, $1.4$). The difference in ejected dust mass exhibits a marked decline with age owing to the Kroupa IMF being richer in AGB stars. 

\subsubsection{Contributions from nebular emission}
\label{subsec:methods_nebular_emission}

Nebular emission from \HII\ regions neighbouring stellar populations can make a significant contribution to the UV continuum slope (see \S \ref{subsec:methods_uv_slope}) and rest-frame far-UV luminosity of galaxies \citep{cullen2024, katz2025nebular, wang2024}. Note that the COLIBRE model explicitly includes \HII\ regions as per \cite{ploeckinger2025} but this implementation assumes a Chabrier IMF. We add an IMF-dependent estimate of nebular continuum and line emission to the SEDs of stellar populations derived from CLOUDY \citep{cloudy}, using tables included with FSPS. We assume that \HII\ regions share the metallicity of their neighbouring stellar populations, as also assumed by \cite{byler2017} who describe the FSPS nebular emission model. We adopt a fixed ratio of ionising hydrogen photons to total hydrogen density of $\log_{10} U=-2$ (the default value in FSPS), though we note that in a self-consistent approach, this ratio would likely be influenced by variations in the IMF. We find that, while the inclusion of nebular emission has a significant impact on far-UV luminosity, varying $\log_{10} U$ from $-4$ to $-1$ has negligible impact on the far-UV luminosity of simulated galaxies, though it does affect the shape of the spectrum. Example far-UV spectra for several values of the ionisation parameter $U$ are shown in Appendix \ref{appendix:nebular_emission}. We find that accounting for the nebular emission of a Chabrier IMF stellar population of metallicity $Z=0.03\,\mathrm{Z}_\odot$ increases its far-UV luminosity by a factor of $\simeq1.4$ at an age of 1 Myr.

\subsection{Dust attenuation modelling}
\label{subsec:methods_dust}

We calculate the influence of dust-attenuation on the SEDs and far-UV luminosities of galaxies using the radiative transfer code SKIRT \citep{skirt}, which models the scattering and absorption interactions between photons and dust, accounting for the size and chemical composition of dust grains. SKIRT utilises the dust composition and size distribution predicted by the live dust model in COLIBRE, which is detailed by \cite{trayford2025} and uses COLIBRE's chemodynamics model \citep{correa2026}. 

COLIBRE yields good agreement with observations of the $z=0$ galaxy dust mass function, $z=0$ dust scaling relations and the cosmic evolution of dust density up to $z\leq5$ \citep{trayford2025}, as well as the observed UV attenuation and submillimetre emission at $z=0$ \citep{lu2026a}. We note, however, that the dust model is not well constrained at earlier epochs, and that the optical properties of the dust grains derive from local observations. Moreover, accurate modelling of the internal structure of molecular clouds requires significantly higher resolution than the simulations used here. Nevertheless, simulations that do not include a live dust model must approximate dust-to-gas mass ratios in post-processing from metal abundances, and make strong assumptions about the spatial distribution of the dust, which erodes their predictive power. The choice of IMF influences the formation and evolution of dust grains and the metals that they accrete (e.g. as shown in Figs. \ref{fig:imf_boosts} and \ref{fig:time_evolved_ssp}), so a self-consistent live dust model is a beneficial aspect of the COLIBRE simulations for our purposes.

A comprehensive description of the method for coupling COLIBRE's live dust model to SKIRT is provided by \cite{gebek2026}, so we present only a brief summary here, highlighting in particular where our approach differs. Although it was not calibrated to reproduce particular observations, the COLIBRE-SKIRT pipeline broadly reproduces the cosmic SED at $z=0$. As detailed by \cite{trayford2025}, the dust model tracks three dust compositions: carbonaceous (graphite), magnesium-silicate (forsterite), and iron-silicate (fayalite), with each species tracked at two grain sizes, $0.01\,\micron$ and $0.1\,\micron$, yielding a total of six tracked grain types. \textsc{SKIRT} then converts the discrete grain sizes into a continuous distribution. In common with \cite{lu2026b}, we assume a single random viewing angle and apply SKIRT calculations to a galaxy-centred volume of comoving side length $L=257\,\ckpc$ (this specific length derives from a proper side length of 100 pkpc at $z=0$ and scaling with the gravitational softening length at higher redshift, see Table \ref{tab:sims}). This volume is sufficient to encompass all dust and star particles at these redshifts.

Our approach differs from the fiducial COLIBRE-SKIRT method of \cite{gebek2026}, also used by \citet{lu2026b}, in several ways. Firstly, the fiducial approach uses BPASS v2.2.1 \citep{bpass1,bpass2} to model the intrinsic emission properties of stellar populations prior to modelling attenuation and re-emission due to dust. As discussed in \S\ref{subsec:methods_fsps}, we instead use FSPS as it affords greater flexibility in the form of the IMF, but in doing so we omit the effects of binary stars. Binary stars likely have a more significant influence on the emission properties of stellar populations with a top-heavy IMF due to the high binary fraction of massive stars. We use FSPS for both the variable IMF and fiducial simulations in this work to avoid introducing systematic differences. Secondly, the fiducial approach models the emission of young stars (age $< 10\,\Myr$) with the dust-free TODDLERS library \citep{toddlers}, which currently assumes a \citet{chabrier} IMF and models nebular line emission from putative \HII\ regions with a self-consistent approach to calculating the appropriate ionisation parameter. TODDLERS also includes nebular continuum emission, but not in the dust-free mode that is used for the fiducial COLIBRE-SKIRT approach, meaning that it likely underestimates the far-UV luminosities of young galaxies (by e.g. $\simeq0.3$ mags for $M_{\rm UV}<-17$ galaxies at $z=12$, see Appendix \ref{appendix:nebular_emission}). Here, as detailed in \S\ref{subsec:methods_nebular_emission}, we model the nebular emission contribution using CLOUDY (assuming a fixed ionisation parameter) accounting for both nebular line and continuum emission.

Thirdly, the fiducial approach (using the TODDLERS library) models contributions from young ($<10$ Myr) stellar populations assuming a constant (10 Myr-averaged) star formation rate based on the star forming gas in the galaxy. This is to mitigate the effect of stochasticity on the predicted emission, stemming from finite sampling of the star formation history over the prior $10\,\Myr$. Here we prefer to treat stellar populations of all ages as SSPs, in order to retain full self-consistency with the transfer of energy, heavy elements and dust from young stellar populations: recall from Fig. \ref{fig:imf_boosts} that the mass of dust ejected by CCSNe in the first $10\,\Myr$ of a stellar population's lifetime can be up to a factor of 3 greater for a top-heavy IMF ($\alpha = -1.6$) than for a Kroupa IMF. In addition, the appropriate time period over which a constant SFH should be assumed in order to to mitigate the effect of stochasticity is uncertain, but is particularly influential when modelling emission with top-heavy IMFs, further discouraging the use of SFH resampling in this work. 

This difference in how young stellar populations are modelled is the primary cause of the difference in the far-UV emission properties of galaxies in the fiducial simulation between our work and that of \cite{lu2026b}, who present far-UV luminosity functions at $z\geq7$. We quantify and discuss the scale of the effect on our key results that would be induced by adopting the star formation history resampling method in Appendix \ref{appendix:resampling}, but remark here that the resampling approach generally reduces the characteristic far-UV luminosity of galaxies at fixed space density, with the effect being more pronounced for galaxies dominated by top-heavy populations. At $z=12$, galaxies with space density $10^{-4}\,{\rm mag}^{-1}\,\cMpc^{-3}$ are predicted to be $0.6$ magnitudes fainter in the far-UV when applying the star formation history resampling method to our simulation using a variable IMF, whilst the difference is negligible for the fiducial COLIBRE simulation.

We present rest-frame far-UV luminosities for all galaxies (i.e. all identified subhaloes with $M_\star>0$) in snapshots corresponding to $z=(5,7,9,10,12,15)$. For all galaxies that have a non-zero dust mass and intrinsic far-UV magnitude brighter than $M_{\rm UV} = - 15$, we calculate intrinsic and dust attenuated far-UV luminosities using SKIRT. For the remaining galaxies we compute intrinsic far-UV luminosities by aggregating the SEDs of their constituent stellar particles using \textsc{SOAP}. For the SKIRT calculations we launch $10^5$ photon packets per galaxy in the wavelength range of $\lambda = 1250-2800$ \AA. We compute the relative error, $R$, on each SKIRT-simulated galaxy flux using the method described by \cite{camps2018} and, if necessary, increase the number of photon packets until the error decreases below $R=0.1$. We find that $> 98$ percent of galaxies at $z=5$ in both simulations achieve $R<0.1$ without the need to use more than $10^5$ photon packets.

Galaxies with intrinsic brightness fainter than $M_{\rm UV} = - 15$ exhibit a very high space density, but they are not critical for the purposes of this study due to the lack of observational measurements in this regime. We therefore do not process such galaxies with SKIRT, thus significantly reducing computational expense. We assume such galaxies exhibit negligible dust attenuation and equate their observed far-UV luminosity to their intrinsic far-UV luminosity, an assumption we show is justified in \S\ref{subsec:results_deltaMuv}. For reference, we apply SKIRT to $102,473$ galaxies in the variable IMF simulation ($4.7$ percent of all galaxies) at $z=5$, and $874$ ($41.0$ percent) at $z=15$. We evaluated the potential discrepancy in using \textsc{SOAP} compared to running dust-free SKIRT and find that, of the sample of $75,927$ ($102,473$) galaxies at $z=5$ from the fiducial (variable IMF) simulation that we processed with SKIRT, the median discrepancy in intrinsic UV magnitude between these methods is $0.0013$ ($0.0014$) mags, with $99.99$ ($99.97$) percent falling within an error of 0.05 mags.

\begin{figure}
    \centering
    \includegraphics[width=\columnwidth]{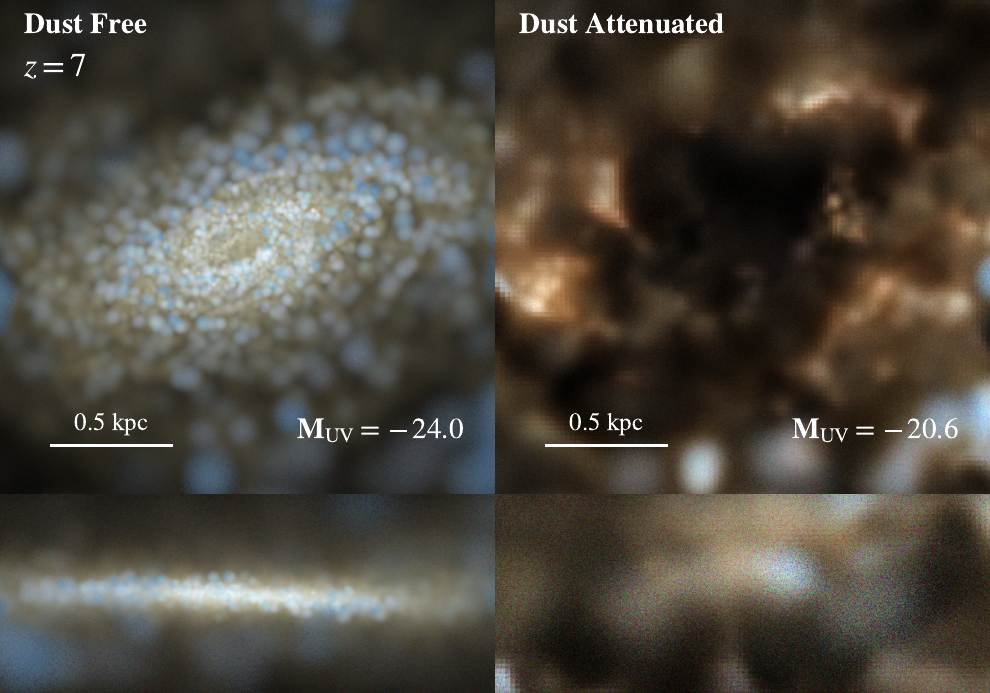}
    \caption{Face-on (\textit{top}) and edge-on (\textit{bottom}) mock images using \textit{JWST} NIRCam filters of the most massive galaxy at $z=7$ in the fiducial COLIBRE L100m6 simulation, $M_{\rm star}=3.8\times 10^{10} \rm \,\Msun$, SFR $=125.5 \rm \,\Msun$yr$^{-1}$, $M_{\rm dust}=8.0\times 10^{7} \rm \,\Msun$, $R_{1/2,{\rm star}}=0.35\,\pkpc$. The images were created by post-processing the simulated galaxy  using the radiative transfer code SKIRT. The left panels shows the galaxy in the absence of dust, right panels include the attenuation effect of dust, which decreases the galaxy's UV brightness by 3.4 magnitudes. RGB colour channels correspond to \textit{JWST} NIRCam Broadband filters (F200W, F150W, F115W) which, for $z=7$ sources, have central rest-frame wavelengths ($2485$, $1876$, $1443$) \AA, respectively.}
    \label{fig:skirt_image}
\end{figure}

To give a visual impression of the emission properties of a COLIBRE-simulated galaxy post-processed with SKIRT, Fig. \ref{fig:skirt_image} shows mock images of the most massive galaxy at $z=7$ in the fiducial COLIBRE L100m6 simulation, in face-on (\textit{top}) and edge-on (\textit{bottom}) orientations, generated by SKIRT without (\textit{left}) and with (\textit{right}) dust attenuation. The RGB colour channels correspond to \textit{JWST} NIRCam broadband filters F200W, F150W and F115W respectively, spanning rest-frame wavelengths $\simeq1200-2900$ \AA$\,$ (far-UV to near-UV) at $z=7$. Following \cite{gebek2026}, we define the smoothing length of stellar particles as the distance to the 32$^{\rm nd}$ nearest stellar particle. The galaxy has a current star formation rate $\rm{SFR} = 125.5\,\Msun\,{\rm yr}^{-1}$, stellar mass $M_\star = 3.8\times 10^{10}\,\Msun$, dust mass $M_{\rm dust} = 8.0\times 10^{7}\,\Msun$, and average dust surface density $\Sigma_{\rm dust} = 1.0 \times 10^8 \,\Msun\,\kpc^{-2}$. The high dust surface density strongly obscures rest-frame UV light, particularly at the centre of the galaxy, from which the emission is almost entirely extinguished. In the rest-frame far-UV ($\simeq1500$ \AA) the attenuation corresponds to $3.4$ magnitudes (based on a single random viewing angle), reducing the far-UV magnitude from the intrinsic value of $M_{\rm UV}=-24.0$ to an observed value of $M_{\rm UV}=-20.6$. This level of attenuation is common for massive, high-redshift galaxies in both the variable IMF and fiducial simulations, as we show in \S\ref{subsec:results_dust}.

\subsection{Characterising the UV continuum slope}
\label{subsec:methods_uv_slope}

The UV continuum slope, $\beta$, is a useful measure of a galaxy's dust attenuation \citep[e.g.][]{calzetti1994} and is sensitive to nebular emission \citep{cullen2024}. We compute both intrinsic and dust attenuated $\beta$ by fitting (with the least absolute deviation method) a power law of the form $f_{\rm\lambda} \propto \lambda^{\beta}$ to composite intrinsic and attenuated galaxy spectra in the wavelength range $1250-2600$ \AA. We find that nebular line emission has a negligible effect on $\beta$, but the choice of ionisation parameter $U$ affects the nebular continuum in this wavelength range such that stronger ionisation yields slightly redder slopes for dust-free stellar populations. We show in Appendix \ref{appendix:nebular_emission} that neglecting nebular emission can artificially steepen dust-free $\beta$ values significantly.

\section{Model validation}
\label{sec:validation}
\begin{figure}
    \centering
    \includegraphics[width=0.9\columnwidth]{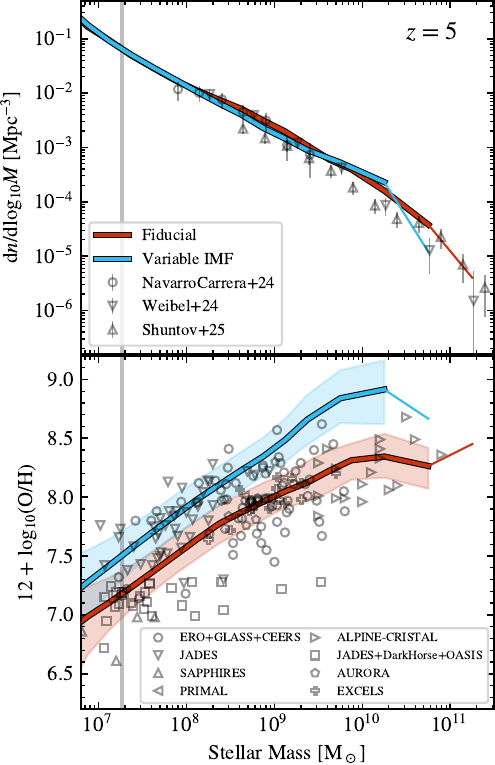}
    \caption{(\textit{Top}) Galaxy stellar mass function and (\textit{bottom}) median gas-phase oxygen abundance as a function of stellar mass, at $z=5$, of the fiducial and variable IMF simulations (see Table \ref{tab:sims}). Note that stellar masses in the vIMF model cannot be directly compared with the observations, which assume a Solar neighbourhood IMF. In the lower panel, only star-forming galaxies (specific SFR $> 10^{-1}\,\Gyr^{-1}$) are shown, with curves denoting median values and shaded regions indicating the $10^{\rm th}$ to $90^{\rm th}$ percentile scatter. Thin lines are used where stellar mass bins are sampled by fewer than 10 galaxies. The vertical grey line denotes the mass scale corresponding to $10\times$ the baryonic particle mass. Symbols represent measurements inferred from \textit{JWST} observations by \protect\cite{navarro2024,weibel2024,shuntov2025} in the top panel and the compilation by \protect\cite{sharda2026} of metallicities inferred from \textit{JWST} observations in the lower panel. Solar metallicity corresponds to $12+\mathrm{log}_{10}(\mathrm{O/H})=8.69$ \citep{asplund2009}.}
    \label{fig:validation}
\end{figure}

Prior to presenting the main comparisons with high-redshift observables (which follow in \S\ref{sec:results}), we first validate the plausibility of the adopted variable IMF model by examining two diagnostic quantities at $z=5$ (the redshift at which the variable IMF simulation was stopped) that we deem necessary for the variable IMF simulation to reproduce.

As discussed in \S \ref{subsec:methods_imf_parameters}, the principal aim of the parametrisation of the variable IMF model was to inject a similar CCSN energy per unit stellar mass formed as the fiducial model, by replacing the fiducial model's pressure-dependent energy per CCSN with a fixed energy per CCSN and a density-dependent IMF. The ratio of total cumulative CCSN energy injected by $z=(15,10,5)$ from the variable IMF and fiducial simulations is $(0.88,0.95,1.3)$. We can also estimate the difference in injected energy that arises solely from CCSN energetics, by applying our variable IMF model to the birth densities of stellar populations in the fiducial simulation, thus eliminating stochastic differences between the two simulations and the downstream effects of self-consistently modelling IMF variations, which become more significant over time. In this case, the energy ratio by $z=(15,10,5)$ is $(1.0,0.98,1.1)$, demonstrating that the 30 percent increase in energy injection in the variable IMF model by $z=5$ is primarily driven by physical effects that stem from the adoption of top-heavy IMF, such as increased heavy element and dust yields. We note that there is likely also a non-negligible effect from driving variations with density instead of thermal pressure. 

Since we aim to maintain feedback energetics in the variable IMF simulation that are comparable to those of the fiducial simulation, a natural validation test for the variable IMF simulation is reproduction of the $z=5$ GSMF produced by the fiducial simulation. The $z=5$ GSMF of the two simulations are shown in the upper panel of Fig. \ref{fig:validation}, demonstrating that they are very similar and deviate in stellar mass at fixed space density by no more than $\simeq0.3$ dex. This verifies that the small differences in CCSN feedback energy injection between the two simulations does not lead to the emergence of galaxy populations with markedly different masses at this epoch. The variable IMF simulation does not produce galaxies as massive as the fiducial simulation, primarily because an SSP with a top-heavy IMF loses more mass via stellar evolution, and does so more rapidly, than is the case for a Solar neighbourhood IMF (see Fig. \ref{fig:time_evolved_ssp}). \citet{chaikin2025} demonstrate that the fiducial simulation accurately reproduces the GSMF inferred from $z=5$ \textit{JWST} observations under the assumption of a Solar neighbourhood IMF, and we include these data on Fig. \ref{fig:validation} for completeness, but caution that these cannot be compared directly with the GSMF of the variable IMF simulation. Had we wished to calibrate the variable IMF simulation to observational data at $z=5$, the most appropriate choice would be the observed rest-frame optical luminosity function. We provide a comparison of the fiducial and variable IMF simulations to the rest-frame optical luminosity function at $z=5$ in Appendix \ref{appendix:z5_optical_diagnostics}. 

For a fixed initial stellar mass, a stellar population with a top-heavy IMF will synthesise a greater mass of oxygen than one with a Solar neighbourhood IMF, making the mass-metallicity relation a critical validation diagnostic \citep[e.g.][]{barber1}. The lower panel of Fig. \ref{fig:validation} therefore shows the mass-metallicity relation, i.e. the median oxygen abundance of the ISM as a function of galaxy stellar mass, of galaxies at $z=5$ in the variable IMF and fiducial simulations. We follow \citet{colibre1} by defining the ISM as gas with density $n_{\rm H}>0.1\,\percc$ and temperature $T<10^{4.5}\,\K$, and by computing the oxygen abundance as the ratio of the total number of gas-phase oxygen nuclei, thus excluding oxygen depleted onto dust grains, to the total number of hydrogen nuclei. We also consider only star-forming galaxies, which at $z=5$ we define as those with a specific SFR $ > 10^{-1}\,\Gyr^{-1}$ (roughly $0.1/t_{\rm H}$, where $t_{\rm H}$ is the Hubble time). At fixed stellar mass, galaxies indeed exhibit an elevated oxygen abundance in the variable IMF simulation relative to counterparts in the fiducial simulation: at $M_\star = 10^9\,\Msun$ the median metal mass fraction is a factor $\simeq 2$ greater. A similar elevation was reported by \cite{cueto2024}, who find a factor $\simeq 1.6$ increase at the same mass scale (albeit at $z=6$) relative to a version of their semi-analytic model adopting a Salpeter IMF. They also find minimal evolution in the mass-metallicity relation at higher redshifts. Fig. \ref{fig:validation} also shows metallicities inferred from \textit{JWST} observations (ERO+GLASS+CEERS; \citealt{nakajima2023}, JADES; \citealt{curti2024}, SAPPHIRES; \citealt{hsiao2025}, PRIMAL; \citealt{sarkar2025},  ALPINE-CRISTAL; \citealt{faisst2026}, JADES+DarkHorse+OASIS; \citealt{isobe2026}, AURORA; \citealt{sanders2026}, EXCELS; \citealt{stanton2026}), as compiled by \citet{sharda2026}, who present a detailed analysis of the mass-metallicity relation of the fiducial COLIBRE simulation at $0\leq z \leq15$. Crucially, both simulations are broadly consistent with the observations for $M_\star \lesssim 10^{10}\,\Msun$, a regime for which the scatter at fixed stellar mass that is comparable to the difference in median oxygen abundance between the two simulations. At higher masses, observations from the ALPINE-CRISTAL survey are more readily reconciled with the mass-metallicity relation of the fiducial simulation.

Clearly, consistency with observations at $z=5$ is no guarantee of consistency at later epochs: the fiducial simulation has been shown to reproduce the $z=0$ mass-metallicity relation well \citep[see Fig.\,20 of][]{colibre1} and, if the factor $\simeq 2$ offset between the simulations were to persist to the present day, the variable IMF model would be challenging to reconcile with the metallicity of galaxies in the local universe, though we note that this level of offset is comparable to well-known systematic uncertainties relating to the calibration of metallicity indicators. The mass-metallicity relation at intermediate- to low-redshift is therefore likely to be amongst the strongest constraints on the degree to which the star formation history of massive galaxies can be dominated by top-heavy stellar populations.

\FloatBarrier
\section{Results}
\label{sec:results}

\begin{figure*}
    \centering
    \includegraphics[width=2\columnwidth]{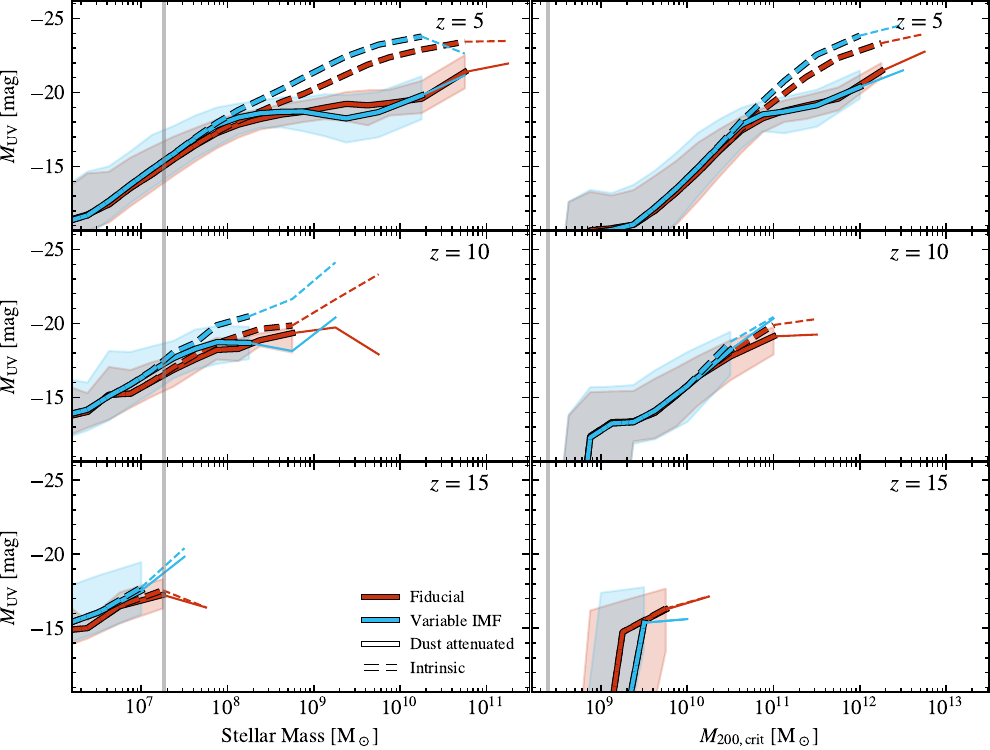}
    \caption{The far-UV magnitude of central galaxies as a function of their stellar mass (\textit{left}) and halo mass (\textit{right}), in the fiducial (red curves) and variable IMF (cyan curves) simulations. The top, middle and bottom panels show the relation at $z=5,10,15$ respectively. Solid curves represent median dust-attenuated UV magnitudes, dashed curves represent dust-free (intrinsic) UV magnitudes, and thin lines are used where mass bins are sampled by fewer than 10 galaxies. Shaded regions show the $10^{\rm th}$ to $90^{\rm th}$ percentile scatter. Vertical grey lines are drawn at the mass scale corresponding to $10\times$ the initial baryonic particle mass (\textit{left}) and $100\times$ the CDM particle mass (\textit{right}). The inclusion of satellite galaxies in the left panels makes no noticeable change. }
    \label{fig:Muv}
\end{figure*}

In this section we present results from the variable IMF simulation, and compare them to observational measurements, and to results from the fiducial simulation. A full analysis of the $z\geq7$ UV luminosity functions of the COLIBRE simulations and related properties is presented by \cite{lu2026b}. Key information about the two simulations are provided in Table \ref{tab:sims}. We consider all galaxies identified in each simulation at each redshift, where galaxies are identified from all subhaloes with non-zero stellar mass, with the exception of Fig. \ref{fig:Muv} which is limited to galaxies in central haloes. Where relevant, observational measurements have been adjusted to adopt the same dimensionless Hubble constant as in the simulations, $h=0.681$. Throughout, thinner lines are used in figures to denote where bins are sampled by fewer than 10 simulated galaxies. Where relevant, we bin the simulated galaxy magnitudes into bins of $\Delta M_{\rm UV} = 0.5$ for $M_{\mathrm{UV}}>-20$, and $\Delta M_{\rm UV} = 1.0$ for $M_{\mathrm{UV}}<-20$. We bin galaxy stellar and halo masses by 0.25 dex, but broaden this to 0.5 dex at high mass ($>10^{10}\,\Msun$ at $z=5$). Binned quantities are plotted at the centre of the corresponding bin.

The section is structured as follows: \S\ref{subsec:results_Muv}, \S\ref{subsec:results_uvlf} and \S\ref{subsec:results_uv_density} present the far-UV magnitude scaling relations, far-UV luminosity functions, and cosmic far-UV luminosity density at redshifts $5\leq z \leq15$, respectively, showing both intrinsic and dust-attenuated far-UV fluxes. \S\ref{subsec:results_dust} examines the influence of the IMF on the dust properties of the simulated galaxy population.

\subsection{UV magnitude scaling relations}
\label{subsec:results_Muv}

Fig. \ref{fig:Muv} shows the median dust-free (intrinsic; dashed curves) and dust-attenuated (solid curves) rest-frame far-UV AB magnitudes of central galaxies as a function of their stellar mass (left) and their halo mass (right) at $z=(5,10,15)$ in the fiducial (red curves) and variable IMF (cyan curves) simulations. We note that including satellite galaxies in the left panel does not affect the results due to the low abundance of satellite galaxies at these mass ranges. We show the halo mass ($M_{200,\mathrm{crit}}$) scaling relation because comparisons of simulations adopting different IMFs at fixed stellar masses, even when using the same initial conditions, do not necessarily compare the same galaxies, because of the impact of the IMF on the mass evolution of individual stellar populations \citep[e.g.][]{woodrum2024}, and because of the divergent galaxy evolution stemming from differing feedback energetics and metal yields. 

As shown by the bottom panel, at $z=15$ the most massive galaxies in either simulation have mass $M_\star < 10^8\,\Msun$, making their star formation histories poorly sampled at m6 resolution. The mass of the most-massive galaxies is also artificially limited by the finite simulation volume, which cannot sample large-scale modes in the power spectrum. To give a sense of the scale of this limitation: the masses of the most-massive galaxies in the fiducial COLIBRE L100m7 and L400m7 simulations at $z=15$ are $4\times10^7\,\Msun$ and $3\times10^8\,\Msun$, respectively. Despite the low masses of galaxies in the L100m6 simulations at this epoch, they are bright: galaxies of $M_\star \sim 10^7\,\Msun$ typically have $M_{\rm UV} \simeq (-17.0,-17.5)$ in the (fiducial, variable IMF) simulation. At $z=15$, the amount of dust attenuation in the UV is relatively low ($<1$ mag for the most massive, and intrinsically brightest, galaxies). 

Broadly, the intrinsic rest-frame far-UV luminosity of galaxies is a monotonic function of their stellar mass in both simulations at all epochs, though there is a significant scatter about the median relation at all stellar masses, which is largely driven by the characteristic age of the galaxy, with younger galaxies being brighter. Though the $10^{\rm th}$ to $90^{\rm th}$ percentile scatter is only shown for the attenuated UV magnitudes, the scatter in the intrinsic UV magnitudes is similar; at $z=10$ we find a scatter in the attenuated $M_{\rm UV}$ of $1.3$ ($1.8$) mags at $M_\star = 10^8\Msun$ compared to $1.6$ ($1.8$) mags for intrinsic $M_{\rm UV}$ in the fiducial (variable IMF) simulations. The median intrinsic luminosity at fixed stellar mass declines with advancing cosmic time, in a similar fashion to the decline of the median specific SFR \citep[shown for the fiducial COLIBRE simulations in Fig. 7 of][]{chaikin2025}. Similar evolution of the $M_{\rm UV}-M_{\star}$ relation is seen in semi-analytic simulations adopting variable top-heavy IMFs conducted with the ASTRAEUS \citep{hutter2025} and GALFORM \citep{lu2025} models. 

The right panels of Fig. \ref{fig:Muv} show there is also a monotonic relationship between rest-frame far-UV luminosity and halo mass, for haloes above the mass scale for which haloes typically host a luminous galaxy ($M_{200,\rm crit}\gtrsim 10^9 \,\Msun$). In both simulations, galaxies at fixed halo mass are brighter at earlier times, in a similar fashion to the trend at fixed stellar mass. This provides a simple means of explaining the trend: the physical size of dark matter haloes at fixed mass is smaller at earlier times, such that the characteristic density of star-forming gas is higher, leading to a higher characteristic star formation rate.

The median intrinsic far-UV luminosity of the brightest galaxies at fixed redshift tends to be greater (at fixed stellar mass) in the variable IMF simulation, owing to the ongoing formation of stellar populations with a top-heavy IMF. At $z=10$, galaxies of mass $M_\star = 10^8\,\Msun$ have median intrinsic far-UV brightness of $M_{\rm UV} \simeq (-19, -20)$ in the (fiducial, variable IMF) simulation. The sensitivity of the median brightness to the IMF at $z=5$ is clear from comparison of the dashed red (fiducial) and cyan (variable IMF) curves in the top panels of Fig. \ref{fig:Muv}, highlighting an offset in intrinsic brightness of $\simeq 1\,{\rm mag}$ for galaxies of $10^9\,\lesssim M_\star /\Msun \lesssim 10^{10.5}$. In our model, the IMF becomes more top-heavy (flatter high-mass slope) for star particles formed at high natal gas density. In Appendix \ref{appendix:alpha_evolution} we discuss that, for $z>7$, the median high-mass IMF slope declines (becomes less top-heavy) with decreasing redshift due to the evolution of the mean cosmic density. For $z<7$, however, star formation with a top-heavy IMF becomes more abundant due to advancing structure formation leading to larger overdensities. The latter is responsible for the offset in intrinsic UV magnitudes between the two models at $z=5$. 

Dust attenuation is significant for galaxies of $M_{\rm UV} \lesssim -17$; an expected, yet interesting, outcome is that the additional dust and metal production by the elevated number of CCSNe of top-heavy stellar populations results in attenuation also being stronger at fixed stellar mass for galaxies in the variable IMF simulation: at $z=5$, galaxies of stellar mass $10^{10}\,\Msun$ exhibit $\simeq (3, 4)$ magnitudes of attenuation in the (fiducial, variable IMF) simulation, compensating (or even over-compensating) the greater intrinsic UV brightness, such that dust-attenuated UV brightnesses at fixed stellar mass in the variable IMF simulation are similar to (or even fainter than) those in the fiducial simulation. We also note that the amount of dust attenuation in the UV luminosities of the most massive galaxies is higher at lower redshift, with a $\simeq4$ mag ($\simeq2$ mag) drop in UV magnitude at $z=5$ ($z=10$) in the variable IMF simulation. This is driven by the mass of the most massive galaxies increasing with decreasing redshift. At these epochs dust attenuation scales with galaxy stellar mass due to a strong correlation between dust surface density and stellar mass, which we examine in detail in \S\ref{subsec:results_dust}. 

\subsection{UV luminosity function}
\label{subsec:results_uvlf}

\begin{figure*}
    \centering
    \includegraphics[width=1.8\columnwidth]{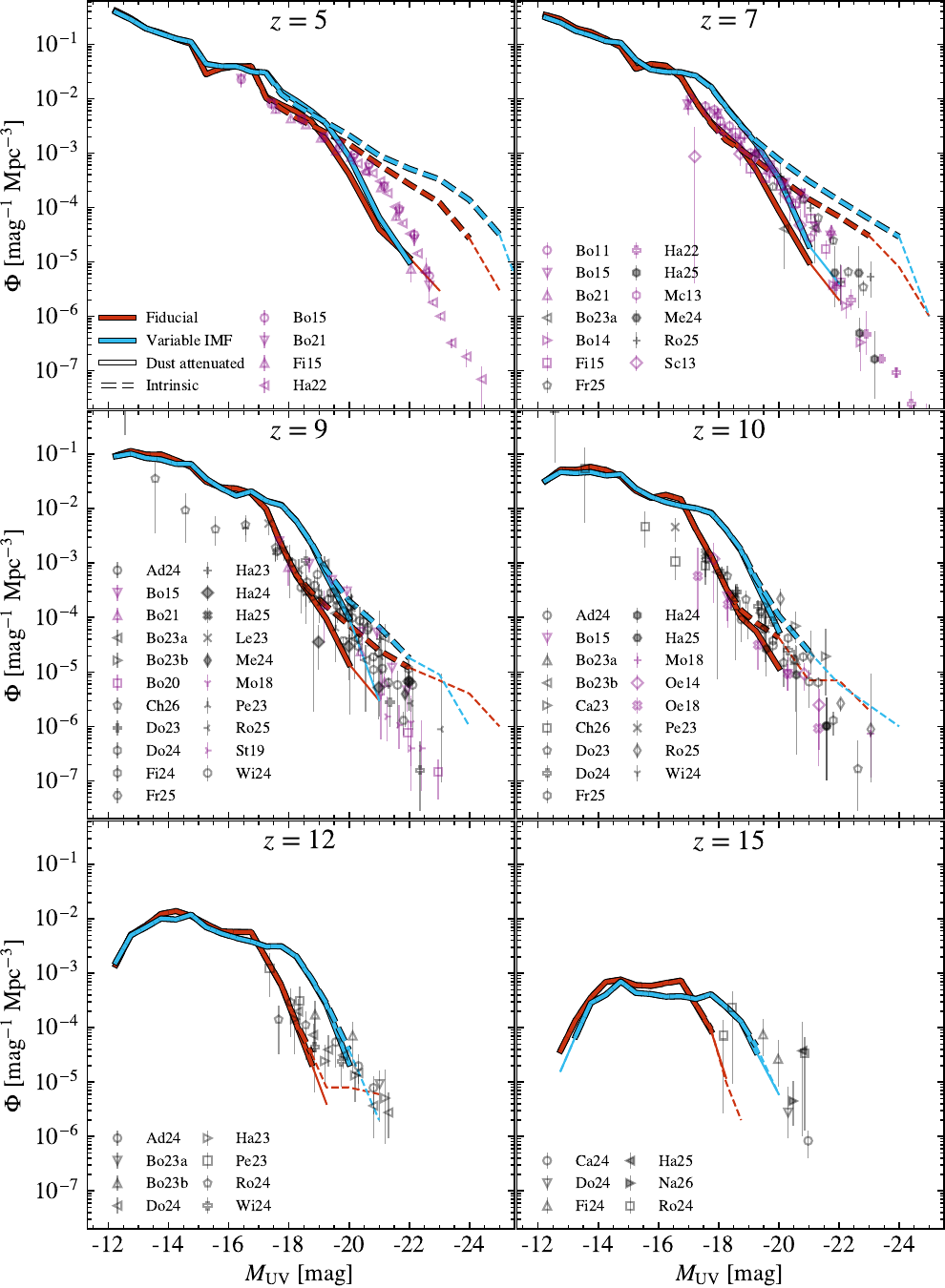}
    \caption{
    Evolution of the dust-attenuated (solid curves) and intrinsic (dashed curves) rest-frame far-UV luminosity function (UVLF) of the L100m6 fiducial COLIBRE (red curves) and variable IMF (cyan curves) simulations from $z = 5$ (top left) to $z = 15$ (bottom right). Thin lines indicate where magnitude bins are sampled by fewer than 10 galaxies. Black and purple symbols represent observational data, detailed in \S\ref{subsubsec:uvlf_data}.}
    \label{fig:uvlf}
\end{figure*}

In this subsection we compare the UV luminosity functions of the fiducial and variable IMF simulations to one another, and to observational measurements. We remind the reader that \citet{lu2026b} demonstrate that the fiducial COLIBRE simulations underestimate the $z>10$ UVLF when accounting for dust attenuation, but note that the UVLF we recover here from this simulation differs slightly from that presented by \citet{lu2026b} primarily because, as discussed in \S\ref{subsec:methods_fsps}, we do not resample the star formation history of young stellar populations. 

The UVLFs of the fiducial (red curves) and variable IMF (cyan curves) simulations are shown in Fig. \ref{fig:uvlf} at redshifts $5\leq z\leq15$. Solid lines denote the UVLF accounting for dust attenuation while dashed lines denote the dust-free UVLF (i.e using intrinsic UV magnitudes). We remind the reader that, as discussed in \S\ref{subsec:methods_dust}, we assume no attenuation for galaxies intrinsically fainter than $M_{\rm UV} = -15$, or if the galaxy has zero dust mass, to avoid unnecessary, computationally-expensive SKIRT calculations. That the UVLFs derived from the intrinsic and dust-attenuated UV magnitudes converge at $M_{\rm UV} \simeq -17$ at all redshifts indicates that this is a conservative choice.

\subsubsection{Observational data}
\label{subsubsec:uvlf_data}

We compare to the following observational datasets: (i) \textit{JWST} data with spectroscopic redshifts \citep{harikane2024, harikane2025, meyer2024, naidu2026}, denoted by filled black symbols; (ii) JWST data based on photometric redshifts \citep{adams2024, bouwens2023a, bouwens2023b, casey2024, castellano2023, chemerynska2026, donnan2023a, donnan2024, finkelstein2024, franco2025, harikane2023, leung2023, perezgonzalez2023, robertson2024, rojasruiz2025, willott2024}, denoted by open black symbols; and iii) pre-JWST data with photometric redshifts \citep{bouwens2011b, bouwens2015, bouwens2021, bowler2014, bowler2020, finkelstein2015, harikane2022, mclure2013, morishita2018, oesch2014, oesch2018, schenker2013, stefanon2019}, denoted by open purple symbols. For clarity, we exclude data points that are only upper limits. We further exclude datasets with redshift bin width $\Delta z > 2$ to ensure an accurate comparison between the simulations and the observational measurements. Following \cite{harikane2025} and \cite{lu2026b}, we omit $M_{\rm UV}>-21$ data by \cite{meyer2024} owing to their sample being incomplete at these magnitudes.

\subsubsection{Considerations when using the UVLF as a constraining diagnostic}

Several factors should be borne in mind when comparing the simulated and observed UVLFs. Firstly, binning observed galaxies into redshift ranges generally biases the population toward the lower-redshift end of the bin \citep{page2000}. \citet[][see their Fig. 8]{lu2026b} examine the impact of this bias on the interpretation of the UVLF of the fiducial COLIBRE simulations at $z\geq7$. For the maximum redshift bin size used in this work, $\Delta z=2$, they find that a redshift-binning bias could artificially boost the UV brightness at fixed space density by at most $\simeq0.5$ mags at $z\geq12$, but this is not enough to alleviate the tension between the fiducial COLIBRE model and observations. Secondly, we must consider the impact of Eddington bias \citep{eddington1913}, i.e. that random errors on the measured luminosities result in a greater number of relatively faint galaxies being `scattered' into high-luminosity bins than the converse, yielding a systematic overestimate of the space density of the brightest galaxies. The impact of Eddington bias on the $z\geq8$ GSMF of the fiducial COLIBRE simulations is shown by \citet{chaikin2025}, and on the $z\geq7$ UVLF by \cite{lu2026b}. For $7\leq z \leq15$, \cite{lu2026b} find that introducing a random Gaussian error of $0.5$ magnitudes to the brightness of simulated galaxies increases the inferred brightness of galaxies with a space density of $10^{-6}\,{\rm mag}^{-1}\,\cMpc^{-3}$ by $\simeq 0.5$ magnitude, though we note that the error on observed magnitudes in the sources shown in Fig. \ref{fig:uvlf} is estimated to be $0.1-0.2$, so the effect of Eddington bias is expected to be less significant than implied by this example (as also concluded by \citealt{lu2026b}). 

Thirdly, we do not account for the effect of cosmic variance on the observational data that may arise from the relatively small cosmic volumes probed by high-$z$ \textit{JWST} surveys. For example, \cite{yung2024} found that a 100 arcmin${}^2$ field can lead to an error in the estimated number densities of galaxies of $\simeq20-30$ percent at $z = 11$ or up to $80$ percent at $z=14$; this is a comparable, though generally larger, field size than those used for the data listed in this section (e.g. \citealt{harikane2023} combines 4 fields for a total area of $\simeq90$ arcmin${}^2$). Fourthly, for the majority of the data compared to in this section, galaxy redshifts were derived from photometric data only. As stated, spectroscopically confirmed data is indicated by filled markers, but this only applies to four of the datasets shown \citep{harikane2024, harikane2025, meyer2024, naidu2026}. \cite{lu2026b} find that introducing a fractional photo-$z$ error of $\sigma_z/(1+z)\geq0.1$ is required to explain the tension between $z>10$ observed UVLFs and the fiducial COLIBRE model (when accounting for dust attenuation), though this exceeds the estimated median error of the data ($0.05$, excluding catastrophic errors). Finally, we remark that AGN are assumed to contribute negligibly but nebular emission makes a significant contribution to the observed UV luminosities of simulated galaxies at $z\geq10$. Fig. \ref{fig:uvlf_nebem} in Appendix \ref{appendix:nebular_emission} shows that the bright end of the UVLF of both the fiducial and variable IMF simulations ($M_{\rm UV}<-17$ and $<-19$ respectively) at $z=15$ shifts faintward by $\simeq0.5$ mags when nebular emission is omitted. Example spectra of stellar populations with a comparable metallicity to that of typical $z=12$ galaxies in either simulation ($Z=0.03\,\rm{Z}_\odot$), including and omitting nebular emission, are also shown in Fig. \ref{fig:nebular_emission}.

\subsubsection{Comparing the fiducial model to observations}

Despite our choice to not resample the star formation history of young star particles (which, as we show in Appendix \ref{appendix:resampling}, typically reduces the UV magnitude at fixed space density around the `knee' of the UVLF, by e.g. $\simeq1$ magnitude for $\Phi\sim10^{-3}\,{\rm mag}^{-1}\,\cMpc^{-3}$ at $z=12$), we find that the fiducial simulation typically underestimates observations of the bright end ($M_{\rm UV}<-18$) of the UVLF at $z>10$, as concluded by \cite{lu2026b}. The same conclusion is reached at $z=15$, where all but the brightest galaxies in the fiducial simulation have very little dust. The limited volume of this simulation means we cannot compare to the most luminous observations at this epoch, with space densities $<10^{-5}\,{\rm mag}^{-1}\,\cMpc^{-3}$, e.g. JADES-GS-z14-0 \citep[$M_{\rm UV} = -20.81 \pm 0.16$, $z=14.32_{-0.20}^{+0.08}$,][]{carniani2024} and MoM-z14 \citep[$M_{\rm UV} = -20.23 \pm 0.06$, $z=14.44 \pm 0.02$,][]{naidu2026}. 

At $z=12$ the fiducial simulation yields a small number of galaxies intrinsically brighter than $M_{\rm UV}=-20$, but their significant dust masses result in the UVLF at a space density of $10^{-5}\,{\rm mag}^{-1}\,\cMpc^{-3}$ being at least a magnitude fainter than inferred from observations. The rarity of galaxies with intrinsic brightness of $M_{\rm UV} < -19$ indicates that appealing to a reduced dust formation efficiency, a longer dust formation timescale, or more efficient dust removal \citep[e.g.][]{ferrara2023, fiore2023, shen2023, ziparo2023, yung2024}, would not resolve the apparent tension between the observations and the simulation, as also found by \cite{lu2026b}. At $z=10$, such solutions would alleviate the tension significantly, as the predicted characteristic intrinsic brightness of galaxies at fixed space density are consistent with (or even brighter than) the observationally-inferred UVLF. However, if one takes the dust content predicted by the simulation, and its effect on attenuation as modelled by SKIRT, at face value, then the intrinsically-brightest galaxies are already heavily obscured at $z=10$, resulting in a significant steepening of the UVLF and leaving no galaxies with observed brightness $M_{\rm UV} < -20$. For $z<10$ the fiducial simulation yields UVLFs that are broadly consistent with the observations, except for a small ($<1$ magnitude) shortfall in brightness for space densities $<10^{-3}\,{\rm mag}^{-1}\,\cMpc^{-3}$ at $z=5-7$. At a space density of $10^{-4}\,{\rm mag}^{-1}\,\cMpc^{-3}$, the characteristic attenuation is $\simeq2.7$ magnitudes at $z=5$, indicating that a mild reduction of the dust surface density would bring the simulations into excellent agreement with the observations. 

\subsubsection{Comparing the fiducial and variable IMF simulations}

At $z=15$, the characteristic brightness of galaxies with space density $\lesssim 10^{-4}\,{\rm mag}^{-1}\,\cMpc^{-3}$ is $\simeq 1.3$ magnitudes brighter in the variable IMF simulation (whether dust attenuation is included or not), illustrating the significant influence on the UVLF that the form of the IMF can induce. We caution, however, that comparison with photometrically-derived observations at this epoch is uncertain due to the possibility of interlopers. The brightest galaxies in the variable IMF simulation have intrinsic UV magnitudes of $M_{\rm UV} \simeq -20.5$, comparable to MoM-z14 and JADES-GS-z14-0, but even at this epoch these rare, early-forming simulated galaxies have significant dust masses and surface densities ($M_{\rm dust} \sim 10^4\,\Msun$; $\Sigma_{\rm dust} \sim 10^3 \,\Msun\,\kpc^{-2}$), yielding $\simeq 0.5$ magnitudes of attenuation. This is a result of the prompt ejection of large masses of dust by top-heavy stellar populations; the median high mass IMF slope in these intrinsically bright ($M_{\rm UV} \simeq -20$) galaxies at $z=15$ is $-1.65$ which results in a factor of $3.0$ times more dust mass ejected from CCSNe by $10\,\Myr$ than is the case for a population with a Solar neighbourhood IMF (see Fig. \ref{fig:imf_boosts}). 

Whilst the predicted attenuation of bright galaxies in the simulation precludes accurate reproduction of the bright end of the observed UVLF at $z=15$, a mild increase of the intrinsic brightness of bright galaxies (e.g. from a greater fraction of their stellar populations being born with a top-heavy IMF), or a mild reduction of their attenuation (e.g. from a reduced dust yield from CCSNe, enhanced dust destruction from CCSNe reverse shocks, or more efficient dust removal in outflows) would bring the simulations into good agreement with the observations at space densities of $10^{-5} \lesssim \Phi/({\rm mag}^{-1}\,\cMpc^{-3}) \lesssim 10^{-4}$. The brightest observed galaxies at this epoch have inferred brightnesses $M_{\rm UV} \simeq -21$: even if one assumes, contrary to what is predicted by the simulation, that dust plays no role at this early epoch, this is a magnitude brighter than the brightest galaxy in the variable IMF simulation. We note, however, that the simulation is inhibited from producing extreme sources by its relatively small volume. We show in Appendix \ref{appendix:convergence} that a volume only 8$\times$ larger (i.e. $L=200\,\cMpc$) is predicted to yield galaxies with intrinsic brightness up to $M_{\rm UV} \simeq -23$ at $z=15$ using the variable IMF model, and would be in good agreement with even the brightest observational constraints on the UVLF at this redshift. This suggests that it is not necessary to invoke more extreme top-heavy star IMFs, nor preclude the presence of dust, to reproduce the brightest sources observed at $z=15$.

At $z=12$ the variable IMF simulation is in good agreement with the bright end ($\Phi < 10^{-4}\,{\rm mag}^{-1}\,\cMpc^{-3}$) of the observationally-inferred UVLF. It is particularly interesting, however, that already at this epoch, more common galaxies in the variable IMF simulation are too bright: at space densities of $10^{-4} < \Phi/({\rm mag}^{-1}\,\cMpc^{-3}) < 10^{-3}$, galaxies are roughly 1 magnitude brighter than observed. The over-brightness of this population of galaxies gradually declines and by $z=5$ the fiducial and variable IMF simulations yield similar UVLFs. This shortcoming of the model highlights that changes to the IMF can readily create fresh tensions between the simulations and observational data. Adopting a higher pivot density ($n_{\rm H,pivot}$ in Eq. \ref{eqn:sigmoid}), so that fewer stellar populations are born with a top-heavy IMF, would alleviate this particular tension, but without re-running the simulation it is not clear whether this simple change would yield a less realistic UVLF on other scales, or at earlier times. We also find that using the same SFH resampling method as in \cite{lu2026b}, which involves replacing the SFH over the last 10 Myr with an average value over that time (see \S \ref{subsec:methods_dust}), would alleviate the tension at $z=9-12$. However, this would also reduce the intrinsic UV magnitude of the brightest galaxies at $z=15$ by $\simeq1$ magnitude. Modifications to the star formation history and the effect on the UVLF are explored in Appendix \ref{appendix:resampling}. 

The (dust-attenuated) UVLF of the variable IMF simulation steadily converges towards that of the fiducial simulation towards $z=5$, in a similar fashion to the convergence of the $z=5$ GSMFs of the two simulations (see Fig. \ref{fig:validation}). For low-mass galaxies this stems in part from the (well-motivated) choice to adopt a Solar neighbourhood IMF for stellar populations that form from relatively low density gas: the characteristic natal density of stellar populations declines monotonically with cosmic time until $z\simeq 7$ (owing to the cosmic expansion, discussed in Appendix \ref{appendix:alpha_evolution}), resulting in a declining fraction of stellar populations being born with top-heavy IMFs. However, the near-convergence of the dust-attenuated UVLFs is not mirrored by the UVLFs derived from intrinsic brightnesses, indicating that the galaxies in the variable IMF simulation that do experience top-heavy star formation arrive at broadly similar $z=5$ far-UV luminosities as their counterparts in the fiducial simulation by having a greater dust mass (as we show in Fig. \ref{fig:dust_mstar_relations}), compensating their greater intrinsic brightness. At $z=5$, and a space density of $\Phi = 10^{-4}\,{\rm mag}^{-1}\,\cMpc^{-3}$, the difference between the intrinsic and dust-attenuated characteristic brightness is $2.7$ ($3.6$) magnitudes in the fiducial (variable IMF) simulation. At lower space densities, the difference is even more pronounced in the variable IMF simulation; we speculate that this stems from the rapid increase of the typical natal density in massive galaxies that begins at $z \simeq 7$ (shown in Fig. \ref{fig:overdensity} and discussed in Appendix \ref{appendix:alpha_evolution}), resulting in a rapid increase of the formation of stellar populations with top-heavy IMFs in this redshift interval.

Fig. \ref{fig:uvlf} demonstrates that the variable, top-heavy IMF model used here alleviates tension between galaxy formation models and $z \gtrsim 10$ \textit{JWST} observations, by boosting the rest-frame far-UV luminosity of rare galaxies. However, this implementation of a variable IMF also overpredicts the UV brightness of galaxies with a greater space density at $z=9-12$. Exploration of the parameter space of the variable IMF model is required to establish whether the tension with \textit{JWST} observations can be alleviated without introducing fresh shortcomings, which will be carried out in future work.


\subsection{Evolution of the far-UV luminosity density}
\label{subsec:results_uv_density}

\begin{figure}
    \centering
    \includegraphics[width=1\columnwidth]{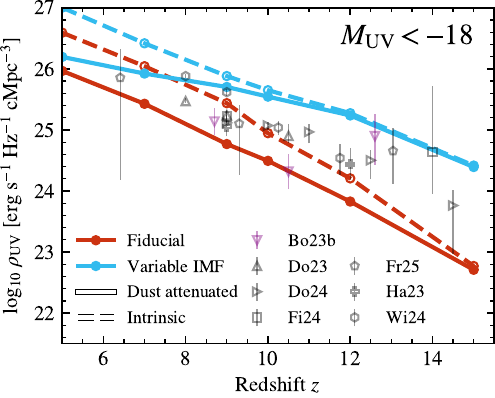}
    \caption{Redshift evolution of the dust-attenuated (solid curves) and intrinsic (dashed curves) UV luminosity density $\rho_{\mathrm{UV}}$ from our L100m6 fiducial (red curves) and variable IMF (cyan curves) simulations, including only galaxies where $M_{\mathrm{UV}}<-18$. Black markers represent observational data where we calculated $\rho_{\mathrm{UV}}$ from published parametric fits to the UVLF in order to use a different magnitude threshold to that in the original study \protect\citep{harikane2023,donnan2023a,donnan2024,finkelstein2024,weibel2024,franco2025}. Purple markers represent data that used a threshold of $-18$ in the original calculation of $\rho_{\mathrm{UV}}$ \protect\citep{bouwens2023b}.}
    \label{fig:uv_density}
\end{figure}

Fig. \ref{fig:uv_density} shows the evolution of the cosmic far-UV luminosity density, $\rho_{\rm UV}(z)$, of the fiducial (red curves) and variable IMF (cyan curves) simulations. As in prior figures, dust-attenuated (intrinsic) UV luminosities are shown using solid (dashed) curves. The UV luminosity density is computed by summing the UV luminosity of all galaxies with magnitudes $M_{\mathrm{UV}}<-18$. While a fainter brightness threshold of $M_{\mathrm{UV}}<-17$ is commonly adopted in observational studies \citep[e.g.][]{donnan2023a, donnan2024, harikane2023}, we restrict our measurement of $\rho_{\rm UV}$ to brighter galaxies. The UVLF is poorly sampled observationally at $M_{\mathrm{UV}}\gtrsim -17$ for $z\gtrsim 10$ (as is clear from Fig. \ref{fig:uvlf}; note that the measurements of \citealt{chemerynska2026} are from strongly-lensed sources), so observational studies that infer the far-UV luminosity density integrated as faint as $M_{\mathrm{UV}} = -17$ at such early epochs generally appeal to faintward extrapolation of functional forms for the UVLF whose parameters are chosen by fitting to lower-redshift observations. \cite{bouwens2023a, bouwens2023b} argue for the use of brighter minimum thresholds for $z \gtrsim 8$ to avoid the introduction of significant additional uncertainty in UV luminosity density estimates that stem from such extrapolation. At $z=10$, the median stellar mass of galaxies with $M_{\rm UV} = -17\pm{0.05}$ is $M_\star\simeq1.8 \times 10^6 \,\Msun$ in both the fiducial and variable IMF simulations, illustrating that galaxies this faint in the two simulations are also, in general, poorly sampled. The observational measurements on Fig. \ref{fig:uv_density} therefore correspond to integration of the UVLF for $M_{\rm UV}<-18$. The measurements of \citet[][purple symbols]{bouwens2023b} adopt this threshold natively, whilst the estimates shown with black symbols correspond to UV luminosity density values that we have computed by integrating a double power-law function over the range $-25\leq M_{\rm UV}\leq -18$, with parameters chosen by fitting the UVLF as per the studies shown in the legend \citep{harikane2023,donnan2023a,donnan2024,finkelstein2024,weibel2024,franco2025}. We make the simple assumption that the fractional error on $\rho_{\rm UV}(M_{\rm UV}<-18)$ is equal to the fractional error on $\rho_{\rm UV}(M_{\rm UV}<-17)$ specified by those studies, as we cannot calculate the true error without access to full details of how the data were fit.

As could be foreseen from inspection of the UVLFs (Fig. \ref{fig:uvlf}), the UV luminosity density is systematically greater in the variable IMF simulation than the fiducial counterpart for $z \ge 5$, during which the cosmic star formation rate density increases monotonically. The difference between the two models declines at later times: at $z = (5,10,15)$ they differ by factors of $(1.7, 11.3, 48.2)$. Comparison with observational measurements indicates that the fiducial simulation generally underestimates observational measurements for $z>7$, as also concluded by \cite{lu2026b}, whilst the variable IMF simulation generally overestimates $\rho_{\rm UV}$ for $9\leq z\leq 12$. This broadly follows from the variable IMF simulation yielding galaxies that are too bright at space densities corresponding to the `knee' of the UVLF, i.e. $10^{-4} \lesssim \Phi/({\rm mag}^{-1}\,\cMpc^{-3}) \lesssim 10^{-3}$ for $9 \lesssim z \lesssim 12$, as such galaxies dominate the star formation rate density \citep[see e.g.][]{chaikin2025}. As noted in \S\ref{subsec:results_uvlf}, the adoption of a higher pivot density ($n_{\rm H, pivot}$ in Eq. \ref{eqn:sigmoid}), would likely reduce the excess brightness of these galaxies and reduce the offset between the variable IMF simulation and observations. At $z>12$ the variable IMF model is consistent with the $\rho_{\rm UV}$ estimates of \citet{bouwens2023b,finkelstein2024} and \citet{franco2025}, but is systematically higher than those of \cite{donnan2024} by $\simeq 0.5\,{\rm dex}$. This highlights that the UV luminosity function is a more informative and robust tool than the UV luminosity density when comparing simulations with observations.

\subsection{Dust and attenuation properties of galaxies}
\label{subsec:results_dust}

\begin{figure}
    \centering
    \includegraphics[width=\columnwidth]{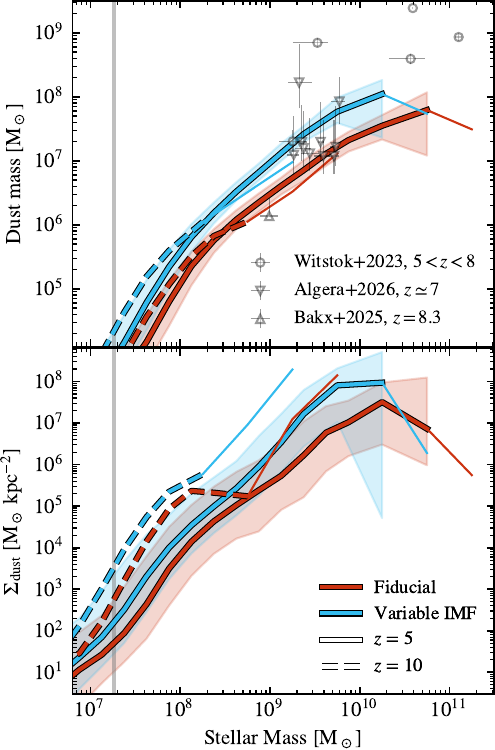}
    \caption{The median galaxy dust mass (\textit{top}) and median galaxy-averaged dust surface density (\textit{bottom}) as a function of galaxy stellar mass, shown for the fiducial (red) and variable IMF (cyan) simulations at $z=5$ (solid curves) and $z=10$ (dashed curves). Note that the inferred stellar mass estimates from observations assume a Solar neighbourhood IMF. Shaded regions denote the $10^{\rm th}$ to $90^{\rm th}$ percentile scatter. Thin lines are used where stellar mass bins are sampled by fewer than 10 galaxies. The vertical grey line indicates the mass scale corresponding to $10\times$ the initial baryonic particle mass. Observationally-inferred dust masses for individual high-$z$ galaxies are denoted by black symbols on the top panel.}
    \label{fig:dust_mstar_relations}
\end{figure}

We turn next to an examination of how a variable, top-heavy IMF impacts the dust properties of galaxies, and the resulting attenuation of their far-UV emission. We showed in Figs. \ref{fig:imf_boosts} \& \ref{fig:time_evolved_ssp} that an SSP formed with a top-heavy IMF yields greater masses of both ejected metals and ejected dust. Moreover, we showed in Fig. \ref{fig:validation} that, at $z=5$, the variable IMF simulation yields galaxies that are significantly more metal rich than similarly-massive counterparts in the fiducial COLIBRE simulation. Besides SSPs with a top-heavy IMF ejecting more dust from CCSNe, an elevated ISM metallicity aids dust grain growth by accretion within the ISM \citep{hirashita2014}. 

\begin{figure*}
    \centering
    \includegraphics[width=2\columnwidth]{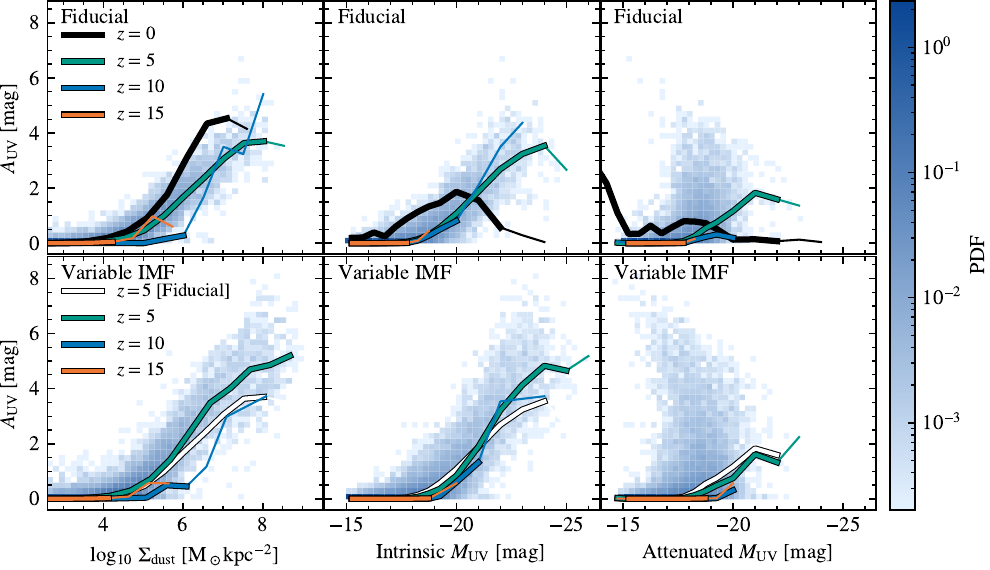}
    \caption{Correlations between the total UV attenuation of galaxies, $A_{\mathrm{UV}}$, and the galaxy-averaged dust surface density (\textit{left}), the intrinsic UV magnitude of galaxies (\textit{middle}), and attenuated UV magnitude of galaxies (\textit{right}), in the fiducial (top row) and variable IMF (bottom row) simulations. Underlying two-dimensional histograms show the probability density of galaxies at $z=5$. Solid curves correspond to the median value at $z=0$ (black, fiducial simulation only), $z=5$ (green), $z=10$ (blue) and $z=15$ (orange). The median relations of the fiducial simulation at $z=5$ are repeated on the lower row in white to aid comparison.}
    \label{fig:delta_Muv}
\end{figure*}

Fig. \ref{fig:dust_mstar_relations} shows the median dust mass (top panel) and dust surface density (bottom panel), as a function of stellar mass, at $z=5$ (solid curves) and $z=10$ (dashed curves), for the fiducial (red) and variable IMF (cyan) simulations. We show the dust surface density as it is more closely connected to the expected attenuation of the far-UV emission. The dust surface density is defined as $\Sigma_{\mathrm{dust}}=M_{\mathrm{dust}} / (2 \pi R_{\rm dust}^2)$, where $R_{\rm dust}$ is the dust half-mass radius. In both simulations, galaxies exhibit a dust mass that is a monotonically-increasing function of stellar mass, and which does not evolve markedly between $z=10$ and $z=5$. Galaxies of mass $M_\star = 10^9\,\Msun$ ($M_\star = 10^{10}\,\Msun$) in the variable IMF simulation at $z=5$ exhibit a median dust mass of  $9.9\times10^6\,\Msun$ ($8.4\times10^7\,\Msun$), which is a factor of $3.1$ ($3.3$) greater than similarly-massive galaxies in the fiducial simulation. As is clear from the figure, this offset is significantly larger than the scatter about the median in either simulation. We compare to dust masses inferred from recent observations from i) a compilation of $5<z<8$ dusty starburst galaxies and submillimetre-detected galaxies from the JINGLE, HERUS and PG surveys \citep{witstok2023}; ii) $z\simeq 7$ UV-selected galaxies from the REBELS survey \citep{algera2026}; and iii) a $z=8.3$ Lyman-break galaxy MACS0416$\_$Y1 \citep{bakx2025}. Note that these stellar masses are inferred using an assumed Solar neighbourhood IMF, so should not be compared directly with the stellar masses from the variable IMF simulation. 

These observations exhibit large scatter in dust mass at fixed stellar mass, reflecting the marked difference in the means by which the galaxies were initially selected: at $M_\star \sim 10^{10}\,\Msun$ the inferred dust masses span two decades. Both simulations are broadly compatible with the REBELS data, but neither simulation yields galaxies with dust masses as high as those inferred for the majority of the dusty star-forming and submillimetre-detected galaxies ($M_{\rm dust}\sim 10^9\,\Msun$), which could be attributed to the limited volume of the simulations. However, owing to how these galaxies were selected, they likely represent the dustiest galaxies in the observed footprint, and can reasonably be treated as the upper limit of the scatter of the galaxy population. We note that reproducing the fluxes of submillimetre-bright star-forming galaxies (with a median redshift of $z\simeq 2$) was the original motivation for adopting a top-heavy IMF in the GALFORM model \citep{baugh2005}.

Greater dust content at fixed stellar mass may introduce additional tension between the variable IMF simulation and observations at later times. \cite{lu2026a} finds good agreement between the fiducial COLIBRE model and observations in the $z=0$ far-IR and submillimetre luminosity functions, indicating that the fiducial model successfully reproduces dust emission at $z=0$. A significant increase of dust masses at $z\ge5$ due to top-heavy star formation may weaken this agreement. 

The lower panel of Fig. \ref{fig:dust_mstar_relations} shows that the surface density of dust, $\Sigma_{\rm dust}$, is also, broadly, a monotonic function of stellar mass, except for the most massive galaxies. At fixed stellar mass, there is a mild evolution of the surface density towards lower values with decreasing redshift, which follows from a steady increase of the characteristic dust half-mass radius. The latter quantity (not plotted for brevity) is similar in both simulations, thus leaving an offset in the dust surface density at fixed stellar mass between the two simulations, which follows directly from the greater dust mass at fixed stellar mass in the variable IMF simulation. This indicates that the elevated intrinsic far-UV luminosity stemming from a top-heavy IMF is partly compensated by the more efficient dust production. We therefore turn to an examination of dust attenuation in the following sub-section.

\subsubsection{Attenuation of far-UV emission}
\label{subsec:results_deltaMuv}

We characterise the attenuation of far-UV emission on a galaxy-by-galaxy basis as the difference between their intrinsic and attenuated UV magnitudes,
\begin{equation}
    A_{\mathrm{UV}} = M_{\mathrm{UV},\mathrm{attenuated}} -  M_{\mathrm{UV},\mathrm{intrinsic}}.
\end{equation}
Fig. \ref{fig:delta_Muv} shows how this attenuation varies as a function of the dust surface density (left column), intrinsic UV luminosity (middle column), and attenuated UV luminosity (right column) of galaxies, in the fiducial (top row) and variable IMF (bottom row) simulations. Solid curves denote the median attenuation at  $z=0$ (black; fiducial simulation only), $z=5$ (green), $z=10$ (blue) and $z=15$ (orange). The $z=5$ median relations for the fiducial simulation are repeated on the panels of the bottom row to aid comparison with the variable IMF simulation, though we caution that direct comparison at fixed $M_{\rm UV}$ does not necessarily compare the attenuation of galaxies with similar stellar mass or halo mass (as is clear from the scaling relations shown in Fig. \ref{fig:Muv}). The probability density of galaxies at $z=5$ is denoted by the underlying two-dimensional histograms.

As expected, attenuation is a broadly monotonic function of dust surface density at all epochs in both simulations. There is a mild positive evolution with decreasing redshift of the attenuation at fixed dust surface density. This reflects the shifting balance towards the more efficiently-attenuating small grain dust (produced by AGB stars, and by the shattering of large grains) as stellar populations evolve; it is also plausible that the more `explosive' individual feedback events at early epochs (due to the greater $f_{\rm E}$ in the fiducial simulation and the greater fraction of top-heavy stellar populations in the variable IMF simulation) leads to a systemic reduction of the attenuation by creating channels through the dust distribution that enable UV photons to escape unimpeded from the ISM \citep[see e.g.][]{trebitsch2017,naidu2020}.

\begin{figure*}
    \centering
    \includegraphics[width=2\columnwidth]{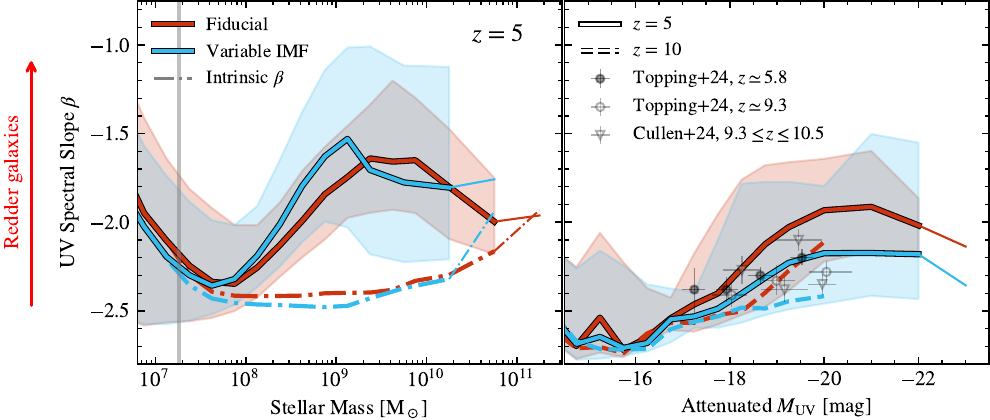}
    \caption{The median UV continuum slope, $\beta$, as a function of stellar mass (\textit{left}) and dust-attenuated UV magnitude (\textit{right}), from the fiducial COLIBRE (red curves) and variable IMF (cyan curves) simulations. In the left panel, solid (dot-dashed) curves correspond to the dust-attenuated (intrinsic) continuum slope. Results are shown at $z=5$ in the left panel, and at $z=5$ (solid curves) and $z=10$ (dashed curves) in the right panel, to highlight the mild redshift evolution. Observational measurements from \protect\cite{cullen2024, topping2024} at $z\simeq 5$ (filled symbols) and $z\simeq 10$ (open symbols) are shown in the right panel. Shading denotes the $10^{\rm th}$ to $90^{\rm th}$ percentile scatter, and thin lines are used where bins are sampled by fewer than $10$ galaxies. The vertical grey line in the left panel shows the mass scale corresponding to $10\times$ the initial baryonic particle mass.}
    \label{fig:beta_slopes}
\end{figure*}

The black curve in the top left panel shows a significant steepening of the $A_{\rm UV}-\Sigma_{\rm dust}$ relation at $z=0$ compared to earlier times, which arises from a markedly greater contribution of small grain dust at $z=0$. For $z>5$, the attenuation at fixed dust surface density is similar in both simulations, indicating that the differing IMF does not significantly influence the dust composition or the galaxy-dust geometry. By $z=5$ however there is a significant offset: at $\Sigma_{\rm dust} = 10^8\,\Msun\,{\rm kpc}^{-2}$, the variable IMF (fiducial) simulation yields a median attenuation of $A_{\rm UV} = 4.8$ ($3.7$). A detailed assessment of the cause of this difference is beyond the scope of this work, but we do not find a significant difference in small-to-large dust grain ratios at fixed dust surface density between the IMF models. We speculate that it is instead a consequence of far-UV emission of bright galaxies in the variable IMF simulation being more centrally-concentrated, which we have found to be true in galaxies with stellar masses $M_\star\geq10^8\,\Msun$ at $z=5$, resulting in greater dust optical depths. We note that only a small contribution to the offset stems from the comparison at fixed $\Sigma_{\rm dust}$ corresponding to galaxies of different stellar mass in the two simulations (see Fig. \ref{fig:dust_mstar_relations}).

The scaling relation connecting the attenuation with intrinsic brightness, for $z\ge 5$, is similar to that connecting the attenuation to the dust surface density, reflecting that both $\Sigma_{\rm dust}$ and the intrinsic $M_{\rm UV}$ scale broadly monotonically with stellar mass, in both simulations. It is interesting to note, however, that in the fiducial simulation at $z=0$, the $A_{\rm UV}-M_{\rm UV}$ relation peaks at $A_{\rm UV} \simeq 2$ for $M_{\rm UV} \simeq -20$, and declines for increasingly bright galaxies, though this region of the relation is poorly sampled.

Inspection of the right panel of Fig. \ref{fig:delta_Muv} highlights that the characteristic attenuation is not single valued at fixed attenuated UV magnitude. Faint observed galaxies can either be intrinsically faint, or heavily attenuated, so measuring the median attenuation at fixed attenuated UV magnitude cannot represent the behaviour accurately. The effect is particularly pronounced in the variable IMF simulation, wherein galaxies of observed $M_{\rm UV} = -16$ can be essentially unobscured, or exhibit up to 8 magnitudes of attenuation. As the attenuation is such a strong function of intrinsic UV brightness at $z=5$, galaxies in both simulations exhibit an effective ceiling in observed UV brightness of $M_{\rm UV} \simeq -21$.

\subsubsection{UV continuum slope}
\label{subsec:results_beta}

The spectral slope of the UV continuum, $\beta$, is a useful diagnostic whose intrinsic value encodes properties of the stellar populations that dominate the emission. The presence of dust reddens the spectrum, affording a route to estimating the dust content of distant galaxies from UV colours. We characterise the UV continuum slope of galaxies as detailed in \S\ref{subsec:methods_nebular_emission}, and reiterate that we model the contribution to the UV spectrum of nebular line and continuum emission. We find the impact of nebular emission on $\beta$ to be significant, with the median (intrinsic) $\beta$ steepening by $0.3$ in the absence of nebular emission for $z=5$ galaxies with intrinsic brightness $M_{\rm UV}\simeq-20$ in both the fiducial and variable IMF simulations (see also Appendix \ref{appendix:nebular_emission}). \cite{cullen2024} similarly find that lowering the contribution from nebular emission yields bluer slopes, such that the minimum expected slope of a stellar population can drop from $\beta \simeq -2.6$ to $\simeq -3.0$ for plausible variations of the assumed ionisation parameter.

The left panel of Fig. \ref{fig:beta_slopes} shows the median UV continuum slope of galaxies in the fiducial and variable IMF simulations as a function of stellar mass at $z=5$. This panel shows both the dust-attenuated (solid curves) and intrinsic (dot-dashed curves) continuum slopes, highlighting the dramatic effect of dust on the measured slope. Results are shown only at $z=5$ as we do not find a strong evolution of $\beta$ at fixed $M_\star$ (but see the right panel). In the absence of dust, both simulations yield (well-resolved, $M_\star \gtrsim 10^8\,\Msun$) galaxies with $\beta \simeq -2.4$, irrespective of mass, with the variable IMF simulation yielding galaxies with marginally bluer spectra.

The inclusion of dust markedly changes the relation, with the median UV slope increasing from the intrinsic value for low mass, largely dust-free galaxies, to a broad peak of $\beta \simeq -1.7$ at $\simeq 10^{9.5}\,\Msun$ in the fiducial simulation, and a sharper peak of $\beta \simeq -1.5$ at $\simeq 10^{9}\,\Msun$ in the variable IMF simulation. A similar turnover in $\beta$ is reported in $z=6.5-7.7$ ALMA observations by \cite{bowler2024} with a peak at $M_{\rm UV}\simeq-21$ to $-22$, which is reasonably consistent with either model. The more efficient formation and growth of dust within galaxies in the variable IMF simulation (see Fig. \ref{fig:dust_mstar_relations}) therefore results in redder UV continuum slopes at $z=5$, despite the presence of stellar populations with top-heavy IMFs. There is a moderate decline in $\beta$ beyond the peak for both simulations, such that massive galaxies ($M_\star\gtrsim 2\times10^9\,\Msun,\,10^9\,\Msun$ for the fiducial and variable IMF simulations respectively) appear slightly bluer. This is slightly offset from the decline in their characteristic dust surface densities, which occurs at $M_\star\gtrsim 10^{10}\,\Msun,\,4\times10^{10}\,\Msun$. This turnover in the UV continuum slope at slightly lower stellar mass is potentially an effect of dust clumping from high levels of CCSN feedback, enabling galaxies with high dust surface densities to appear bluer, and motivates further analysis into this effect.

The right panel of Fig. \ref{fig:beta_slopes} shows the dust-attenuated slope as a function of the dust-attenuated UV magnitude, for both simulations, at $z=10$ (dashed curves) and $z=5$ (solid curves), enabling a comparison with observational measurements and illustrating the weak evolution with redshift. Filled (open) circles denote measurements from \cite{topping2024} for galaxies from redshift bins $5<z<7$ ($8.5<z<11$) with median redshifts $5.8$ ($9.3$), whilst open triangles denote those of \cite{cullen2024} from two redshift bins $7.5\leq z \leq 10$ and $10<z<11$ with collective median redshift $9.9$. We note that $\beta$ is estimated from photometry rather than spectral fitting for both observational samples; \cite{katz2023sphinx} concluded from analysis of synthetic observations of the SPHINX simulations ($4.6 < z < 10$) that estimating $\beta$ from photometry instead of spectral fitting generally steepens the slope by a  median bias of $0.12$ in their sample of $\simeq 14,000$ galaxies. 

The attenuated $\beta-M_{\rm UV}$ relations of the two simulations differ significantly in shape from the attenuated $\beta-M_\star$ relations (left panel, solid curves). The former relations exhibit broader peaks at lower values of $\beta$, particularly for the variable IMF simulation which peaks at $\beta \simeq -2.2$ for $z=5$ in the interval $-22 \lesssim M_{\rm UV} \lesssim -20$ (c.f. $\beta \simeq -1.5$ at $M_\star \sim 10^9\,\Msun$). The difference stems from galaxies of fixed mass exhibiting a wide range of attenuated UV magnitudes (see Fig. \ref{fig:Muv}), which effectively `stretches' the narrow peak of the $\beta-M_\star$ into the broad peak of the $\beta-M_{\rm UV}$ relation. As the range of attenuated UV magnitudes for galaxies with mass $M_\star \sim 10^9\,\Msun$ is greater in the variable IMF simulation than in the fiducial case (owing to galaxies being able to realise greater intrinsic luminosities and greater dust surface densities in the former) the effect is more pronounced in the variable IMF simulation, and results in the median UV continuum slope being bluer at all attenuated $M_{\rm UV}$ in the variable IMF simulation. 

The observations do not strongly discriminate between the simulations: the difference of the median $\beta$ for $-17 \lesssim M_{\rm UV} \lesssim -20$ is comparable to the scatter of the handful of available observational measurements, and fainter galaxies in both simulations are consistent with the data. Galaxies of $M_{\rm UV} \simeq -20$ exhibit slopes that are marginally redder than the data in the fiducial simulation. It is therefore noteworthy that the agreement of the variable IMF simulation with the $z=5$ UVLF (see Fig. \ref{fig:uvlf}), which stems from strong dust attenuation of intrinsically-bright galaxies, is consistent with the observationally-inferred $\beta-M_{\rm UV}$ relation.

\section{Summary}
\label{sec:summary}

We explore whether the formation of stellar populations with top-heavy IMFs (i.e. initially comprising a greater fraction of massive, UV-bright stars than is typical of the Solar neighbourhood) in high-redshift galaxies yields a galaxy population with elevated rest-frame far-UV luminosities. This mechanism has been proposed as a potential solution to the apparent failure of galaxy formation models to yield as many UV-bright galaxies at early epochs as is inferred from \textit{JWST} observations.

We therefore present results from a new cosmological hydrodynamical simulation evolved with a version of the COLIBRE galaxy formation model \citep{colibre1,colibre2} that allows the high-mass slope of the IMF to vary as a function of natal gas conditions. The stellar population yields of heavy elements and dust, and the energetics of their feedback due to CCSNe, are self-consistently adjusted in response to variation of the IMF. We adopt a simple relation between the IMF and natal gas density that results in the injection of a similar feedback energy per unit stellar mass formed as the fiducial COLIBRE simulation. This is achieved by changing the number of CCSNe per unit stellar mass formed (as a natural consequence of varying the IMF), rather than varying the energy injected per CCSN, as is adopted empirically by the fiducial COLIBRE model. This approach therefore provides a possible physical interpretation to variation of the CCSN energy injected per unit mass. Specifically, stellar populations born from relatively low-density gas assume a Solar neighbourhood IMF \citep[that of][]{kroupa}, but those born from denser gas have an IMF whose slope in the high-mass regime ($0.5 < m_\star/\Msun < 100$) is shallower than that of the Kroupa IMF ($\alpha > -2.3$). Stellar populations formed with the maximally top-heavy IMF allowed by the model ($\alpha = -1.6$) initially exhibit far-UV luminosities a factor of $\simeq 4$ greater than those formed with a Kroupa IMF, but also produce roughly twice as many CCSNe, and promptly eject a factor of $\simeq3$ times more metal and dust mass from CCSNe. We evolve a simulation volume of side length $L=100\,\cMpc$ and particle mass $\sim 10^6\,\Msun$ (L100m6) to $z=5$ with this model. We compare the outcomes of this model with the fiducial COLIBRE L100m6 simulation, which has identical initial conditions. In both cases we model the far-UV emission (including nebular radiation) of stellar populations using \textsc{FSPS} \citep{fsps,byler2017}, and model the absorption and scattering of these photons by dust, on a galaxy-by-galaxy basis, by coupling the stellar populations and COLIBRE's live dust model to the \textsc{SKIRT} \citep{skirt} radiative transfer code. 

Our findings are as follows:

\begin{enumerate}
     
    \item By replacing COLIBRE's empirically-motivated relation between the energy injected per CCSN and natal gas conditions \citep[see eq. 2 of][]{colibre2} with an IMF whose top heaviness (and hence number of CCSNe per unit stellar mass formed) is a simple function of natal gas conditions, it is possible to obtain a comparable CCSN energy per unit stellar mass formed with a fixed energy per CCSN (\S\ref{subsec:methods_vimf_model}). We show that this model yields a galaxy population at $z=5$ with a similar galaxy stellar mass function to the fiducial COLIBRE L100m6 simulation  (Fig. \ref{fig:validation}a), enabling the exploration of the influence of a variable IMF on observational diagnostics, relative to a model adopting a universal Solar neighborhood IMF, without the galaxy population being markedly changed by the differing feedback energetics that stem from a top-heavy IMF. Despite galaxies of fixed space density having similar masses in the fiducial and variable IMF simulations, the latter yields galaxies that, at fixed mass, exhibit significantly greater metallicities (Fig. \ref{fig:validation}b).

    \item Galaxies in the variable IMF simulation exhibit greater far-UV intrinsic luminosities at fixed mass than counterparts in the fiducial simulation, for all $z \ge 5$. At $z=10$ galaxies of $M_\star =10^8\,\Msun$ are typically $1$ magnitude brighter in the variable IMF simulation than in the fiducial case. The dust-attenuated luminosities of galaxies in the variable IMF simulation are greater than in the fiducial simulation for $z \gtrsim 7$, but the offset is less than that of the intrinsic luminosity, owing to the elevated dust surface density of galaxies in the variable IMF simulation. By $z=5$, the two simulations exhibit similar attenuated $M_{\rm UV}-M_\star$ relations (Fig. \ref{fig:Muv}). 
    
    \item For $z \gtrsim 7$, the variable IMF simulation reproduces the bright end of the observed far-UV luminosity function (UVLF) more accurately than the fiducial COLIBRE L100m6 simulation, demonstrating that the formation of top-heavy stellar populations at early times can alleviate tension between \textit{JWST} observations and traditional galaxy formation models. At $z=15$ the variable IMF simulation yields galaxies with dust-attenuated (intrinsic) far-UV magnitudes as bright as $M_{\rm UV} \simeq -19.5$ ($M_{\rm UV} \simeq -20$), still short of the inferred far-UV magnitudes of the brightest \textit{JWST}-observed sources at this epoch (Fig. \ref{fig:uvlf}). However our box size convergence test suggests that a volume 8$\times$ larger (i.e. $L=200\,\Mpc$) would yield galaxies with intrinsic brightness $M_{\rm UV} \simeq -23$ at $z=15$ (Fig. \ref{fig:uvlf_big}), indicating that it is unnecessary to invoke more extreme top-heavy IMFs than adopted in the variable IMF simulation, nor to preclude the presence of dust, to reproduce the brightest sources observed at $z = 15$.

    \item For $z \gtrsim 7$, the variable IMF simulation yields galaxies of relatively high space density ($10^{-4} \lesssim \Phi/({\rm mag}^{-1}\,\cMpc^{-3})\lesssim 10^{-2}$), i.e. those at the `knee' of the UVLF, that are roughly half a magnitude too bright. As the fiducial COLIBRE simulation reproduces the UVLF in this regime, the simplest conclusion is that the IMFs of stellar populations forming in these galaxies in the variable IMF model are too extreme, though Appendix \ref{appendix:resampling} demonstrates that modelling recent star formation with a constant star formation history may alleviate this tension. This highlights that changes to the IMF can readily create fresh tensions between the simulations and observational data (Fig. \ref{fig:uvlf}). 
    
    \item At $z=5$, the fiducial and variable IMF simulations reproduce the UVLF with comparable accuracy. This is despite galaxies in the latter exhibiting significantly elevated intrinsic far-UV luminosities. The agreement with the observed UVLF follows from galaxies in the variable IMF simulation exhibiting more attenuation as a result of their greater dust surface densities (Figs. \ref{fig:Muv}, \ref{fig:uvlf} and \ref{fig:dust_mstar_relations}).

    \item The variable IMF simulation yields a higher far-UV luminosity density, $\rho_{\rm UV}$, than the fiducial simulation for all $z \geq 5$. The difference is a factor of $\simeq 48$ at $z=15$, declining to a factor of $\simeq 1.7$ at $z=5$ (including dust attenuation). The elevated UV luminosity density yields better agreement with observational estimates of $\rho_{\rm UV}$ for $z>12$, though the observations exhibit significant scatter. The variable IMF model overpredicts $\rho_{\rm UV}$ at $9\leq z\leq12$ as result of galaxies at the `knee' of the UVLF being too bright (Fig. \ref{fig:uv_density}).
    
    \item In both simulations we find that dust mass is a monotonically-increasing function of stellar mass that does not evolve strongly between $z=10$ and $z=5$, and which exhibits little scatter. At $z=5$ the variable IMF simulation yields dust masses that are a factor of $\simeq 3$ greater than those of the fiducial simulation for galaxies of mass $M_\star = 10^9-10^{10}\,\Msun$, an offset that is significantly greater than the scatter about the median for either simulation. This may have implications for the agreement between the COLIBRE model and $z=0$ observations, such as the FIR and submillimetre luminosity functions presented by \cite{lu2026a}. The dust surface density, which more strongly correlates with far-UV attenuation, is also a broadly monotonic function of stellar mass, and decreases at fixed stellar mass with advancing cosmic time. As the half-mass radius of the dust distribution is similar in both simulations, the dust surface density of galaxies in the variable IMF simulation is higher than in the fiducial simulation, per the dust mass. The elevated far-UV luminosity of galaxies in the variable IMF simulation is therefore partly compensated by their greater dust surface density (Fig. \ref{fig:dust_mstar_relations}).

    \item Dust attenuation is a monotonically-increasing function of the dust surface density, but the slope of the $A_{\rm UV}-\Sigma_{\rm dust}$ relation steepens with advancing time, owing primarily to the growing contribution of small dust grains (ejected from evolved AGB stars, and also resulting from the shattering of large grains). It is also plausible that the efficacy of attenuation at fixed dust surface density (measured on galaxy-wide scales) is reduced at early times by the greater `explosiveness' of feedback events (due to a higher energy per CCSN in the fiducial model, and a higher number of CCSNe per unit stellar mass formed in the variable IMF model) blowing `channels' in the dust distribution, enabling leakage of UV photons. The characteristic attenuation is not single valued at fixed attenuated (i.e. observed) far-UV magnitude: faint observed galaxies can be intrinsically faint, or luminous but heavily attenuated. In the variable IMF simulation, galaxies of observed brightness $M_{\rm UV} = -16$ can be essentially unobscured low-mass galaxies, or massive galaxies with up to 8 magnitudes of attenuation (Fig. \ref{fig:delta_Muv}).

    \item The intrinsic slope of the UV continuum, $\beta$, as a function of stellar mass, is similar for the fiducial and variable IMF simulations, and in both cases is bluer if the contribution to the UV spectrum of nebular emission is neglected. Dust markedly changes the $\beta-M_\star$ relation, with the variable IMF simulation more strongly affected owing to greater characteristic dust surface densities: at $z=5$ the continuum slope in the variable IMF simulation peaks at $\beta \simeq -1.5$ for galaxies of $M_\star \simeq 10^{9}\,\Msun$, whilst in the fiducial simulation it peaks at $\beta \simeq -1.7$ for galaxies of $M_\star \simeq 10^{9.5}\,\Msun$. As galaxies of mass $M_\star \sim 10^9\,\Msun$ exhibit a wide range of attenuated brightnesses, the peaked $\beta-M_\star$ relation translates to a `stretched' $\beta-M_{\rm UV}$ relation, with the variable IMF simulation yielding bluer slopes for all attenuated $M_{\rm UV}$ than the fiducial simulation. Both simulations are broadly consistent with observational measurements of the $\beta-M_{\rm UV}$ relation (Fig. \ref{fig:beta_slopes}).

\end{enumerate}

Our study gives rise to a number of interesting outcomes. Primarily, we conclude that the rest-frame far-UV brightness of $z>10$ galaxies observed with \textit{JWST} can be accommodated by galaxy formation models within the $\Lambda$CDM cosmogony with a plausible adjustment to the assumed IMF at early times, as also concluded by studies using semi-analytic models \citep[e.g.][]{hutter2025,lu2025,fontanot2026}. However, we have also demonstrated that whilst a top-heavy IMF yields intrinsically-brighter stellar populations, these populations yield more CCSNe (assuming other parameters of the IMF are unchanged), eject a significantly greater dust mass, and more strongly enrich the ISM (promoting further dust grain growth). These effects must be self-consistently modelled in order to derive predictions for far-UV luminosities from galaxy formation models. 

The variable IMF simulation analysed here adopts a single, heuristically-motivated parametrisation of how the IMF might vary (i.e. varying the slope of the high-mass regime of the IMF as a function of natal gas density), with the number of CCSNe yielded by the IMF intended to mimic the effect of the calibrated relation between CCSN energetics and natal gas conditions adopted by the fiducial COLIBRE model. It is a success of this model that the variable IMF simulation more accurately reproduces the bright end of the $z \gtrsim 7$ UVLF than the fiducial COLIBRE L100m6 simulation. However the variable IMF simulation also yields galaxies around (or just beyond) the `knee' of the UVLF that are systematically too bright at $7 \le z \lesssim 10$, illustrating how changes to the IMF can readily create fresh tensions between the simulations and observational data. 

\section*{Acknowledgements}

We thank Piyush Sharda and Jonathan Davies for helpful discussions about this work. AD acknowledges an STFC doctoral studentship. RAC and MB acknowledge support from STFC grants ST/Y002482/1 and ST/Y001907/1. CGL acknowledges support from STFC consolidated grants ST/T000244/1 and ST/X001075/1. AG gratefully acknowledges financial support from the Fund for Scientific Research Flanders (FWO-Vlaanderen, project FWO.3F0.2021.0030.01). EC acknowledges support from STFC consolidated grant ST/X001075/1. NA acknowledges financial support  by the Flemish Fund for Scientific Research (FWO-Vlaanderen) through the  research grant G0C4723N. ABL acknowledges support by the Italian Ministry for Universities (MUR) program “Dipartimenti di Eccellenza 2023-2027” within the Centro Bicocca di Cosmologia Quantitativa (BiCoQ), and support by UNIMIB’s Fondo Di Ateneo Quota Competitiva (project 2024-ATEQC-0050). CSF acknowledges support from the European Research Council through Advanced Investigator grant DMIDAS (GA 786910). FH acknowledges funding from the Netherlands Organization for Scientific Research (NWO) through research programme Athena 184.034.002. SP acknowledges support by the Austrian Science Fund (FWF) through grant-DOI: 10.55776/V982. 

This work utilised the Prospero high performance computing facility at Liverpool John Moores University, and the DiRAC@Durham facility managed by the Institute for Computational Cosmology on behalf of the STFC DiRAC HPC Facility (www.dirac.ac.uk). The latter equipment was funded by BEIS capital funding via STFC capital grants ST/K00042X/1, ST/P002293/1, ST/R002371/1, and ST/S002502/1, Durham University and STFC operations grant ST/R000832/1. DiRAC is part of the UK National e-Infrastructure.

\section*{Data Availability}

The data supporting the plots within this article are available on reasonable request to the corresponding author. The COLIBRE simulation code and simulation data will eventually be made publicly available. We intend to include an updated version of the variable IMF code used here with that code release. In the meantime, those interested in using the simulations are encouraged to contact the corresponding author. 



\bibliographystyle{mnras}
\bibliography{references}



\appendix

\section{Box size convergence}
\label{appendix:convergence}

\begin{figure}
    \centering
    \includegraphics[width=\columnwidth]{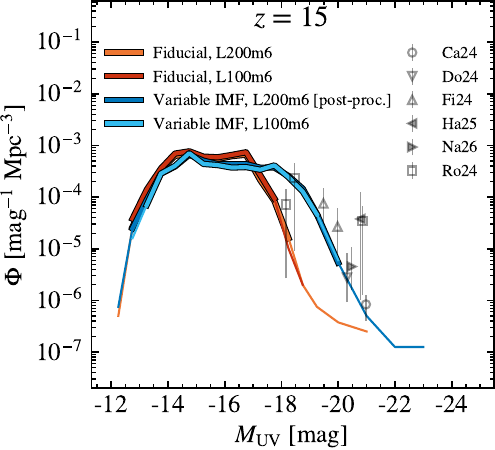}
    \caption{ A box size convergence test of the intrinsic (dust-free) UV luminosity function (UVLF) at $z = 15$, comparing the largest fiducial COLIBRE cosmological volume at m6 resolution (L200m6, (200 cMpc)$^3$ volume), to the simulation volume adopted in this work (L100m6, (100 cMpc)$^3$ volume). We have not run a self-consistent L200m6 variable IMF simulation, so variable IMF galaxy luminosities in this case come from post-processing the fiducial simulation assuming our variable IMF model. We find that the difference between a self-consistent result and post-processing on the UVLF is minimal for $z>10$. Thin lines indicate where magnitude bins are sampled by fewer than $10$ galaxies.}
    \label{fig:uvlf_big}
\end{figure}

We assess the impact of the relatively small volume of the L100m6 fiducial and variable IMF simulations on the bright end of the high redshift UVLF, by comparing the UVLFs of the fiducial L100m6 and L200m6 COLIBRE simulations \citep[both introduced by][]{colibre1}, which have side lengths 100 and 200 cMpc respectively and the same particle mass ($\sim10^6\,\Msun$ for both baryonic and dark matter particles). Fig. \ref{fig:uvlf_big} compares the UVLF at $z=15$ derived from the intrinsic far-UV luminosities (which are good approximations to the observed luminosities of galaxies at this early epoch owing to the minimal influence of dust) of galaxies from the fiducial L100m6 (red curve) and L200m6 (orange curve) simulations. We find good convergence of the UVLF for space densities down to a few times $10^{-6}\,{\rm mag}^{-1}\,\cMpc^{-3}$, corresponding to $M_{\rm UV} \lesssim -18.5$. The UVLF can be probed, albeit with relatively poor sampling, using the L200m6 volume to a few times $10^{-7}\,{\rm mag}^{-1}\,\cMpc^{-3}$, corresponding to $M_{\rm UV} \lesssim -21$ for the fiducial COLIBRE model.

The computational expense of L200m6 simulations precluded the execution of a variable IMF simulation in this larger volume. However, as the self-consistent effects of the top-heavy IMF, relative to the use of a Chabrier IMF, are not yet pronounced by $z=15$, we can approximate the UVLF stemming from the variable IMF model by modelling the emission of stellar populations in the fiducial COLIBRE L200m6 simulation under the assumption that they formed with the IMF specified by equations \ref{eqn:slope_piecewise} and \ref{eqn:sigmoid}. The blue curve therefore shows the UVLF derived from the fiducial L200m6 simulation, with UV luminosities mimicking the variable IMF model, and the cyan curve shows the UVLF derived from the variable IMF L100m6 simulation. The convergence of the cyan and blue curves, in the regime over which the former is well sampled, demonstrates that it is reasonable to `post process' the fiducial simulation in this fashion at this early epoch. We have further tested the convergence of the intrinsic UVLF between self-consistently running and post-processing the variable IMF model with the L100m6 simulations and find good convergence for $5\leq z \leq 15$, owing to the treatment of CCSN feedback in either model, as explained in \S \ref{subsec:methods_imf_parameters}. The larger volume of the L200m6 simulation can therefore be expected to yield galaxies, formed with the variable IMF model, as intrinsically bright as $M_{\rm UV} \simeq -23$, significantly brighter than the most extreme galaxies yet observed by \textit{JWST} at this early epoch. 


\section{IMF high-mass slope values}
\label{appendix:alpha_evolution}

\begin{figure}
    \centering
    \includegraphics[width=\columnwidth]{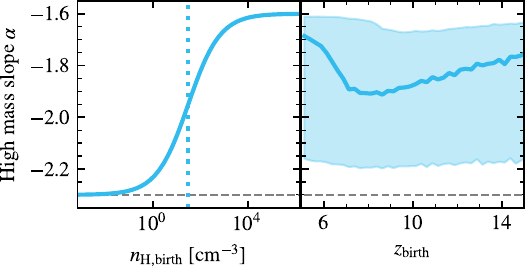}
    \caption{\textit{Left:} The relationship between stellar birth density $n_{\mathrm{H,birth}}$ and the high mass ($m>0.5 \,\Msun$) slope of the IMF (Eq. \ref{eqn:sigmoid}, parameters as described in \S \ref{subsec:methods_imf_parameters}). The horizontal dashed line represents the slope of a Chabrier IMF ($\alpha=-2.3$), while the vertical dotted line represents the pivot birth density, $n_{\rm H,0}=30\,\mathrm{cm}^{-3}$, on which this relation is centred. \textit{Right:} The evolution of the high-mass slope of the IMF for star particles in the L100m6 variable IMF simulation. The curve shows the median slope per birth redshift, while the shaded region denotes the $10^{\rm th}$ to $90^{\rm th}$ percentile scatter.}
    \label{fig:imf_slope}
\end{figure}

\begin{figure}
    \centering
    \includegraphics[width=\columnwidth]{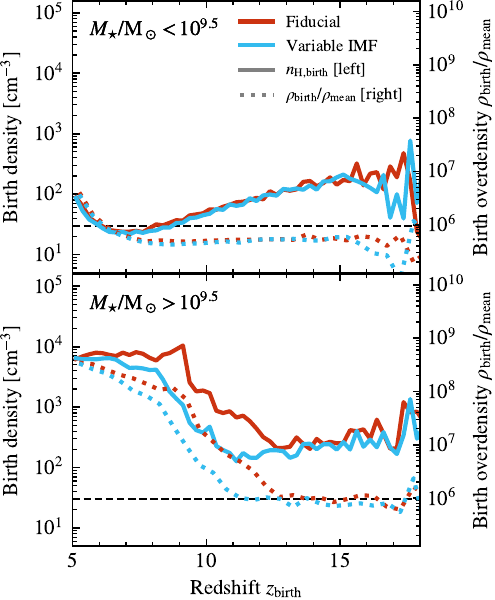}
    \caption{Evolution of the median stellar birth gas density $n_{\mathrm{H,birth}}\equiv\rho_{\mathrm{birth}}/m_{\mathrm H}$ (solid curves, left axis) and stellar birth gas overdensity $\rho_{\mathrm{birth}}/\rho_{\rm mean}$ (dotted curves, right axis) for the fiducial (red) and variable IMF (cyan) simulations. The top (bottom) panel shows the median density of all star particles residing in $M_\star<10^{9.5}\,\Msun$ ($M_\star>10^{9.5}\,\Msun$) galaxies at $z=5$. The overdensity is calculated with respect to the cosmic mean density, $\rho_{\rm mean}$. The horizontal dashed line represents the pivot stellar birth density parameter $n_{\rm H,0}=30\,{\rm cm}^{-3}$ in Eq. \ref{eqn:sigmoid}.}
    \label{fig:overdensity}
\end{figure}

In this appendix we discuss the evolution of the characteristic high-mass slope of the IMF in the variable IMF simulation, which is driven by the evolution of the characteristic stellar birth density. The left panel of Fig. \ref{fig:imf_slope} shows the relation between the high-mass slope of the IMF ($m>0.5\,\Msun$) and stellar birth density $n_{\rm H,birth}$, as described by Eq. \ref{eqn:sigmoid} in \S \ref{subsec:methods_imf_parameters}, used in the variable IMF simulation. The right panel shows the resulting evolution of the high-mass slope for star particles born at $5\leq z_{\mathrm{birth}}\leq15$ in the L100m6 variable IMF simulation. The median slope is relatively top-heavy ($\alpha\simeq-1.75$) at $z=15$ and steepens as the simulation evolves, reaching a minimum of $\alpha\simeq-1.90$ at $z=7$. At later times, however, the slope increases (becomes more top-heavy), reaching a maximum of $\alpha\simeq-1.70$ at $z=5$. Based on the fiducial simulation, which has been run beyond $z=5$, birth densities will decline again for $z<4$.

Fig. \ref{fig:overdensity} shows the evolution of the stellar birth gas density (solid curves, left axis) and the stellar birth gas overdensity (dotted curves, right axis) in two galaxy stellar mass bins (above and below $M_\star=10^{9.5}\,\Msun$ at $z=5$) for the fiducial (red) and variable IMF (cyan) simulations. The horizontal dashed line shows the pivot stellar birth density $n_{\rm H,0}=30\,{\rm cm}^{-3}$ on which the density-slope relation is centred (Eq. \ref{eqn:sigmoid}). The upturn in median IMF slope at $z\simeq7$ is mirrored in the stellar birth densities of galaxies in the low mass bin, while the median birth density of stars born in higher-mass galaxies increases earlier ($z\simeq11$). The latter population is far less abundant, which is why the median IMF slope more closely resembles the density evolution of the low-mass bin. The stellar birth overdensity (dotted lines; values shown on the right axis), defined here as $\rho_{\mathrm{birth}}/\rho_{\rm mean}$ where $\rho_{\rm mean}=\Omega_{\rm b}\rho_{\rm crit}=3\Omega_{\rm b}H(z)^2/(8\pi G)$, is approximately constant for stars formed at $z>8$ ($z>12$) in galaxies in the low-mass (high-mass) bin while the stellar birth density is declining, indicating that the IMF slope is driven by the cosmic expansion of the universe at these times. At later times, formation of massive structures drives up overdensities, causing a rapid upturn in stellar birth densities and, consequently, the high-mass slope of the IMF (Fig. \ref{fig:imf_slope}).


\section{The impact of nebular emission on the UV spectrum}
\label{appendix:nebular_emission}

\begin{figure}
    \centering
    \includegraphics[width=0.9\columnwidth]{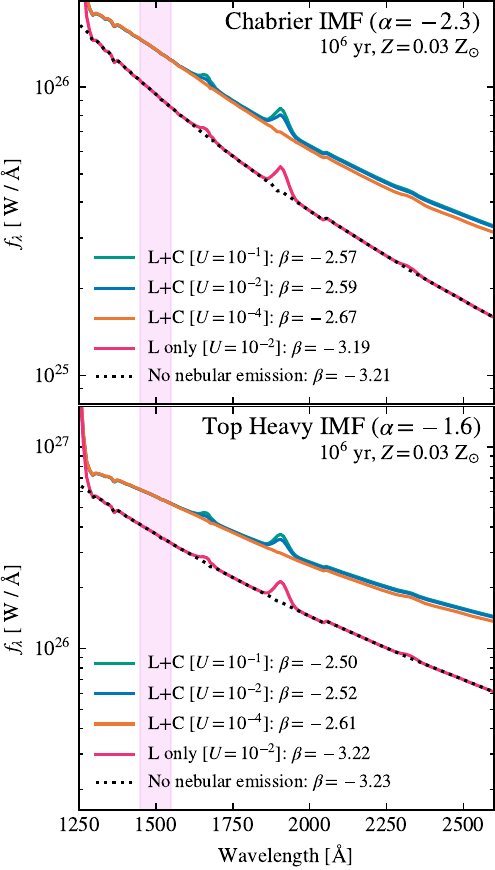}
    \caption{Example spectra ($f_{\mathrm{\lambda}}$) in the wavelength range $1250-2600$ \AA{} for a low-metallicity ($Z=0.03\,\rm{Z}_\odot$) SSP of age $1\,\Myr$ generated with several nebular emission models. The top (bottom) panel shows a Chabrier IMF (top-heavy IMF, $\alpha=-1.6$) SSP. The purple shaded region represents the top hat transmission filter ($1450-1550$ \AA) used for the far-UV luminosity. Each curve corresponds to different nebular emission contributions (L: line emission, C: continuum), showing the ionisation parameter $U$ and the resulting UV continuum slope $\beta$ ($1250-2600$ \AA).}
    \label{fig:nebular_emission}
\end{figure}

As described in \S\ref{subsec:methods_nebular_emission}, we include the contribution of nebular line and continuum emission when modelling the UV spectra of stellar populations, under the assumption that gas producing the nebular emission shares the element abundances of the stellar population. We adopt a fixed ionisation parameter of $U=10^{-2}$. Fig. \ref{fig:nebular_emission} shows the effect of different ionisation parameters on an example SSP UV spectrum with $Z=0.03\,\rm{Z}_\odot$ (assuming $Z_\odot=0.0134$), which is broadly the median metallicity of resolved galaxies ($M_\star\gtrsim 10^7\,\Msun)$ in the fiducial and vIMF simulations at $z=12$. As the contribution of nebular emission declines rapidly for population ages $\gtrsim 3\,\Myr$, we show the spectrum at an age of $1\,\Myr$. The upper panel shows a population with a Chabrier IMF (as adopted by the fiducial simulation) while the lower panel shows that of a maximally top-heavy IMF ($\alpha=-1.6$). `L+C' denotes spectra with nebular line and continuum emission included, while `L only' (pink curve) only includes the nebular lines. For reference we also show the emission from the stellar population in the absence of nebular emission (dotted curve). The shaded purple region represents the $1450-1550$ \AA{ }wavelength window over which we integrate the UV emission (assuming a top hat transmission filter) to estimate far-UV luminosities.

The ionisation parameter $U$ is a dimensionless measure of the intensity of ionizing radiation relative to gas density. We show the UV spectrum with nebular emission for $U=[10^{-4},10^{-2},10^{-1}]$ (green, blue and orange curves, respectively); a higher ionisation parameter yields stronger nebular line emission, most notably the CIII emission line at $1909$ \AA, and mildly reddens the nebular continuum. Whilst increasing $U$ mildly increases the UV continuum slope ($\beta$), the impact over the wavelength range we use to calculate the far-UV luminosity is negligible. The figure also highlights that nebular line emission has negligible influence on the far-UV luminosity and the UV continuum slope. The nebular continuum has a marked effect however: the `L+C' ($U=10^{-2}$) case yields an elevated UV brightness relative to the `No nebular emission' by a factor of $1.4$ ($1.5$) and reddens the slope by  $+0.62$ ($+0.62$) for a $1\,\Msun$ Chabrier IMF (top-heavy IMF) SSP at age 1 Myr. 

\begin{figure}
    \centering
    \includegraphics[width=\columnwidth]{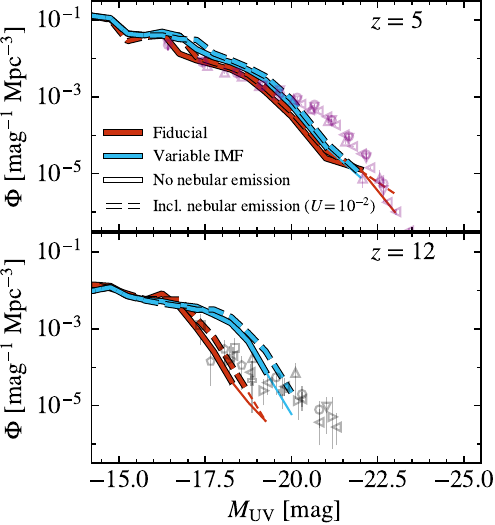}
    \caption{The dust attenuated UV luminosity functions (UVLFs) at $z=5$ (\textit{top}) and $z=12$ (\textit{bottom}) of the fiducial (red curves) and variable IMF (cyan curves) simulations, comparing luminosities accounting for nebular emission (dashed curves, as used throughout this study with an ionisation parameter of $U=10^{-2}$) to those with no nebular emission (solid curves). Thin lines indicate where magnitude bins are sampled by fewer than $10$ galaxies. Black and purple symbols represent observational data, as used in Fig. \ref{fig:uvlf}.}
    \label{fig:uvlf_nebem}
\end{figure}

Fig. \ref{fig:uvlf_nebem} shows the effect of nebular emission on the dust attenuated UVLF at $z=5$ and $z=12$ for both the fiducial (red) and variable IMF (cyan) simulations. Dashed curves show the approach taken in this work, which accounts for nebular (line and continuum) emission with a fixed ionisation parameter of $U=10^{-2}$, while solid curves show the result when only including the stellar continuum (no nebular emission). Omitting nebular emission from our galaxy spectra shifts the bright end of the UV luminosity function at $z=12$ by $\simeq0.5$ mags to fainter UV magnitudes for both IMFs, but the effect diminishes significantly at lower redshift (demonstrated at $z=5$ in the top panel). 


\section{Resampling the formation history of young stellar populations}
\label{appendix:resampling}

\begin{figure}
    \centering
    \includegraphics[width=\columnwidth]{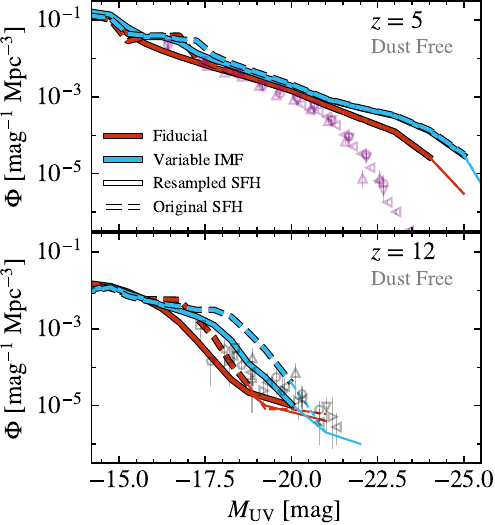}
    \caption{The intrinsic UV luminosity functions (UVLFs) at $z=5$ (\textit{top}) and $z=12$ (\textit{bottom}) of the fiducial (red curves) and variable IMF (cyan curves) simulations, comparing the non-resampled values (`Original SFH', dashed curves, as used throughout this study) to those obtained when replacing the emission of young (age $< 10\,\Myr$) stars with that of a constant star formation history (`Resampled SFH', solid curves) with value set by the time-averaged SFR gas particles in the galaxy. Thin lines indicate where magnitude bins are sampled by fewer than $10$ galaxies. Black and purple symbols represent observational data, as used in Fig. \ref{fig:uvlf}.}
    \label{fig:uvlf_resampled}
\end{figure}

\begin{figure}
    \centering
    \includegraphics[width=0.9\columnwidth]{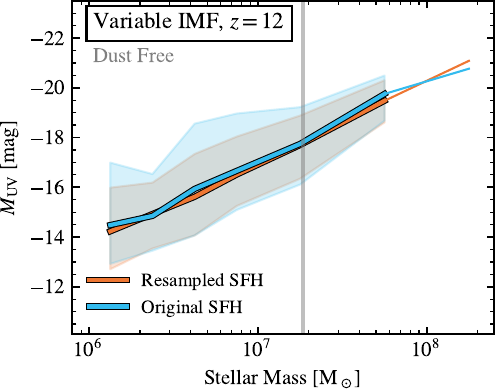}
    \caption{Dust-free UV magnitude - stellar mass relation for the variable IMF simulation at redshift $z=12$, compared for the `Original SFH' and `Resampled SFH' cases described in Appendix \ref{appendix:resampling}. The vertical grey line marks $10\times$ the baryonic particle mass resolution.}
    \label{fig:Muv_resampled}
\end{figure}

As detailed in \S\ref{subsec:methods_dust}, we elect against resampling the star formation history of young stellar populations when computing their far-UV luminosity. The approach used by the fiducial COLIBRE-SKIRT pipeline (\citealt{gebek2026}, as used by \citealt{lu2026b}) replaces the contribution of young (age $<10\,\Myr$) stellar populations with emission corresponding to a constant SFR formation rate, set by the time-averaged ($10\,\Myr$) SFR of gas particles in the galaxy. As discussed in \S\ref{sec:methods}, the modelled UV luminosities of galaxies in the fiducial COLIBRE L100m6 simulations presented here differ from those presented by \cite{lu2026b} primarily because those authors adopt the resampling method. Differences in nebular emission also contribute, as shown in Appendix \ref{appendix:nebular_emission}, but to a lesser extent.

Fig. \ref{fig:uvlf_resampled} shows the effect of this resampling on the intrinsic (i.e. unattenuated) UVLF at $z=5$ and $z=12$, with the approach taken here (`Original SFH', dashed curves) compared against that of the fiducial COLIBRE-SKIRT approach (`Resampled SFH', solid curves). We do not present dust-attenuated resampled luminosities because doing so would require additional SKIRT simulations for each galaxy and Fig. \ref{fig:uvlf} demonstrates that dust attenuation only impacts the observed UV luminosities of the brightest galaxies at $z=12$ ($M_{\mathrm UV}<-19,\,<-20$ for the fiducial and variable IMF simulations respectively). The main effect of resampling the star formation history is a reduction of the scatter in $M_{\rm UV}$ at fixed stellar mass, as shown for the variable IMF simulation at $z=12$ in Fig. \ref{fig:Muv_resampled}. Modelling stellar populations with a 10 Myr-averaged SFH reduces the UV luminosity of galaxies with recent bursts in star formation, suppressing the upscattered (to higher luminosities) population evident from the orange shaded region. This impacts the bright end of the UVLF for $z\geq10$ by reducing the brightness of relatively rare galaxies with space densities $\lesssim10^{-3} {\rm mag}^{-1}\,\cMpc^{-3}$ by $\simeq1$ mag. This would resolve the tension between the variable IMF simulation and \textit{JWST} observations at $z=9-12$ where, as discussed in \S \ref{subsec:results_uvlf}, the variable IMF simulation overpredicts the UV magnitude of galaxies at a space density of $\Phi\sim10^{-3}\,{\rm mag}^{-1}\,\cMpc^{-3}$ by $\simeq1$ mag. However, resampling the SFH would lessen agreement between the variable IMF simulation and observations at $z=15$ by reducing the UV brightness of the galaxies populating the bright end of the UVLF by $\simeq1$ magnitude.


\section{Optical Luminosity Functions}
\label{appendix:z5_optical_diagnostics}

In \S \ref{sec:validation} we verify that the variable IMF model has not significantly modified the galaxy population at $z=5$, the latest redshift that the variable IMF simulation was evolved to, by demonstrating agreement between the GSMF of the fiducial and variable IMF simulations at $z=5$. A shortcoming of this comparison is that the variable IMF simulation cannot be directly compared to the `observed' GSMF, because stellar masses inferred from observations assume a Solar neighbourhood IMF. We therefore supplement that validation test by comparing the two simulations against the corresponding directly-observable property, namely the optical luminosity function. 

\begin{figure}
    \centering
    \includegraphics[width=0.9\columnwidth]{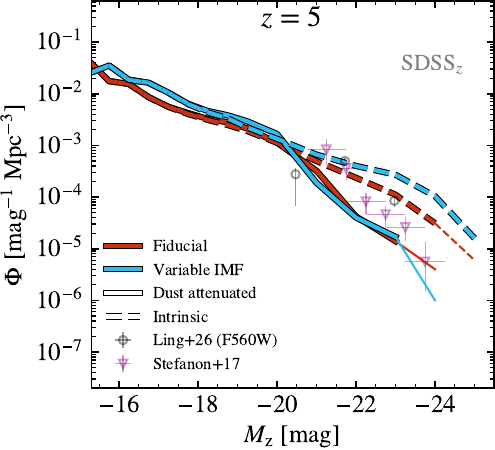}
    \caption{Dust-attenuated (solid curves) and intrinsic (dashed curves) rest-frame z-band ($\lambda\simeq0.9\,\mu \rm{m}$) luminosity function of the L100m6 fiducial COLIBRE (red curves) and variable IMF (cyan curves) simulations at $z = 5$. Thin lines indicate where magnitude bins are sampled by fewer than 10 galaxies. Purple and black symbols represent observational \textit{HST} and \textit{JWST} data by \protect\cite{stefanon2017, ling2026} respectively.}
    \label{fig:optical_lf}
\end{figure}

Fig. \ref{fig:optical_lf} presents the optical ($z$-band) dust-attenuated and intrinsic rest-frame luminosity functions of the fiducial and variable IMF simulations at $z=5$. Optical luminosities were computed with the SDSS $z$-band filter (rest-frame central wavelength $\lambda\simeq\,0.9\mu \rm{m}$) and are compared to data from \textit{HST} \citep{stefanon2017} and \textit{JWST} \citep{ling2026}, where the latter is derived from the observer-frame luminosity function in the \textit{JWST} F560W band ($\lambda_{\rm obs}\simeq5.6\,\mu\rm{m}$). There is good agreement between the dust attenuated $z$-band luminosity functions of the fiducial and variable IMF simulations at $z=5$, as was also the case for the GSMF at this redshift (Fig. \ref{fig:validation}). Both simulations agree within the scatter of the observations at intermediate magnitudes ($-22<M_{\rm z}<-20$) but mildly under-predict the bright end ($M_{\rm z}<-22$) of the luminosity function, by $\simeq0.5$ mags (with the exception of the brightest observed measurement at $M_{\rm z}\simeq-23.7$). Analysis of the luminosity functions, from the far-UV to the submillimetre, produced by the fiducial COLIBRE simulations for $z\leq6$, along with comparison to observations, will be presented in a forthcoming paper by Lu et al. in prep.). 

\bsp	
\label{lastpage}
\end{document}